# iMaNGA: mock MaNGA galaxies based on IllustrisTNG and MaStar SSPs. - III. Stellar metallicity drivers in MaNGA and TNG50

Lorenza Nanni ,[1]★ Justus Neumann ,[2]★ Daniel Thomas ,[1,3] Claudia Maraston,[1] James Trayford ,[1] Christopher C. Lovell ,[1,4] David R. Law,[5] Renbin Yan [6] and Yanping Chen[7]

[1]*Institute of Cosmology & Gravitation, University of Portsmouth, Dennis Sciama Building, Portsmouth PO1 3FX, UK*
[2]*Max-Planck-Institut für Astronomie, Königstuhl 17, 69117 Heidelberg, Germany*
[3]*School of Mathematics and Physics, University of Portsmouth, Lion Gate Building, Portsmouth PO1 3HF, UK*
[4]*Astronomy Centre, University of Sussex, Falmer, Brighton BN1 9QH, UK*
[5]*Space Telescope Science Institute, 3700 San Martin Drive, Baltimore, MD 21218, USA*
[6]*Department of Physics, The Chinese University of Hong Kong, Shatin, N.T., Hong Kong SAR, China*
[7]*New York University Abu Dhabi, Abu Dhabi PO Box 129188, United Arab Emirates*

**ABSTRACT**
The iMaNGA project uses a forward-modelling approach to compare the predictions of cosmological simulations with observations from SDSS-IV/MaNGA. We investigate the dependency of age and metallicity radial gradients on galaxy morphology, stellar mass, stellar surface mass density ($\Sigma_*$), and environment. The key of our analysis is that observational biases affecting the interpretation of MaNGA data are emulated in the theoretical iMaNGA sample. The simulations reproduce the observed global stellar population scaling relations with positive correlations between galaxy mass and age/metallicity quite well and also produce younger stellar populations in late-type in agreement with observations. We do find interesting discrepancies, though, that can inform the physics and further development of the simulations. Ages of spiral galaxies and low-mass ellipticals are overestimated by about 2–4 Gyr. Radial metallicity gradients are steeper in iMaNGA than in MaNGA, a discrepancy most prominent in spiral and lenticular galaxies. Also, the observed steepening of metallicity gradients with increasing galaxy mass is not well matched by the simulations. We find that the theoretical radial profiles of surface mass density $\Sigma_*$ are steeper than in observations except for the most massive galaxies. In both MaNGA and iMaNGA [Z/H] correlates with $\Sigma_*$, however, the simulations systematically predict lower [Z/H] by almost a factor of 2 at any $\Sigma_*$. Most interestingly, for galaxies with stellar mass $\log M_* \leq 10.80\,\mathrm{M_\odot}$, the MaNGA data reveal a *positive correlation* between galaxy radius and [Z/H] at fixed $\Sigma_*$, which is not recovered in iMaNGA. Finally, the dependence on environmental density is negligible in both the theoretical iMaNGA and the observed MaNGA data.

## 1 INTRODUCTION

Current cosmological (magneto)hydrodynamic simulations have the goal of modelling how galaxies form and evolve in the Universe described as from the current cosmological theory and accounting for all the physical processes considered fundamental for galaxies (e.g. the formation of dense cold clouds, stellar formation and evolution, stellar feedback, chemical enrichment, mergers etc.; see for example Schaye et al. 2014; Taylor & Kobayashi 2014; Dolag 2015; McAlpine et al. 2016; Kaviraj et al. 2017; Grand et al. 2019; Nelson et al. 2019a; Feldmann et al. 2023). Their predictions must be tested to evaluate the strength of the current theories on which they are based. Modern galaxy surveys such as the Sloan Digital Sky Surveys (SDSS; York et al. 2000; Abazajian et al. 2003) and the JWST Mission (McElwain et al. 2023; JWST) provide the basis for the direct comparison between the observed and the simulated universe. A powerful way to test the simulations is to virtually observe the simulated galaxies by producing synthetic data products. This forward modelling method has been followed, for example, in Tonini et al. (2010), Torrey et al. (2015), Trayford et al. (2015, 2017), Bottrell et al. (2017), Huertas-Company et al. (2019), Rodriguez-Gomez et al. (2019), Schulz et al. (2020), Sarmiento et al. (2022), and Barrientos Acevedo et al. (2023). With this method, the simulated galaxies are brought into the *observational space*, i.e. synthetic spectra and images are generated.

This approach is at the core of the iMaNGA project (see Nanni et al. 2022, 2023, from now on Paper I and Paper II, respectively). With the first two papers of this series, we presented our methodology to produce synthetic data products. We mimic observations by the survey Mapping Nearby Galaxies at Apache point observatory

★ E-mail: lorenza.nanni@port.ac.uk (LN); jneumann@mpia.de (JN)

(MaNGA; Bundy et al. 2015), an integral field spectroscopy (IFS) survey, by generating MaNGA-like data cubes. In Paper II, we present a method to generate a MaNGA-like sample, constructing the iMaNGA sample, applying the selection criteria for the MaNGA-Primary sample to the cosmological volume. For this project, we work with IllustrisTNG cosmological magnetohydrodynamic simulations (Nelson et al. 2019a), in particular adopting TNG50, which is the simulation with the highest spatial and mass resolution within the IllustrisTNG project. However, the methodology presented in Paper I and Paper II can be applied to any other modern cosmological simulation (e.g. McAlpine et al. 2016; Kaviraj et al. 2017; Davé et al. 2019; Grand et al. 2019; Villaescusa-Navarro et al. 2021; Feldmann et al. 2023).

In this paper, we present a systematic comparison between the theoretical iMaNGA and the observed MaNGA samples. We focus on the direct comparison with work published by our group in Goddard et al. (2017) (hereafter G17) and Neumann et al. (2021) (hereafter, N21). G17 use an early release of the MaNGA galaxy sample to study stellar population properties, focusing on age and metallicity gradients as a function of galaxy mass, type, and environment. N21 investigate the distribution of stellar metallicity within and across galaxies. The paper exploits the complete MaNGA sample, and the stellar population parameters are obtained through full-spectra fitting using the code FIREFLY (Wilkinson et al. 2017). A similar analysis was previously conducted in Lian et al. (2018) on a smaller sample of MaNGA galaxies. The MaNGA sample allows the authors to explore the relation between stellar metallicity and stellar surface mass density both globally and locally. In particular, N21 investigate the relation between stellar metallicity in galaxies as a function of galactocentric radius, and the interplay among stellar metallicity, stellar mass, stellar surface mass density, and galactocentric radius. N21 show that the surface mass density mainly drives the stellar metallicity distribution within galaxies, whereas radial dependencies at fixed surface mass density are secondary.

In this paper, we analyse at the same time both the iMaNGA and the MaNGA sample, presenting the results for both catalogues in the same manner, so that the comparison can be as direct as possible.

The paper is structured as follows: in Section 2, we present the data and models necessary to carry out our analysis. In Section 3, we summarize both Paper I (Section 3.1) and Paper II (Section 3.2) after which we present the iMaNGA Secondary sample in Section 3.3. Furthermore, we discuss the T-morphology and the inclinations for the iMaNGA galaxies (see Section 3.4). In Section 4, we present the global stellar mass–age and –metallicity relations (Section 4.1); the stellar populations' age, metallicity, and stellar surface mass density radial profiles (Section 4.2); the local relation between metallicity and stellar surface mass density (Section 4.3); the local relation between the stellar surface mass density and the effective radius and its trend with stellar metallicity (Section 4.3.3); and finally, the metallicity gradients as a function of galaxy environment (Section 4.4). We draw our conclusions in Section 5.

## 2 DATA & MODELS

Here, we introduce the models and data used in this work. We briefly summarize the MaNGA survey and its main characteristics in Section 2.1. We then give an overview of the IllustrisTNG simulation suites in Section 2.2. The stellar population models used in this work are introduced in Section 2.3. Finally, the FIREFLY MaNGA VAC produced in Neumann et al. (2022) – hereafter, N22VAC – is briefly presented in Section 2.4.

### 2.1 The MaNGA galaxy survey

MaNGA (Mapping Nearby Galaxies at Apache point observatory; Bundy et al. 2015), which is part of the SDSS-IV survey (Blanton et al. 2017) and concluded its observations in 2020 August, observed 10 010 unique galaxies at a median redshift of $z \sim 0.03$, providing the largest sample of IFS data to date (Abdurro'uf et al. 2022).

MaNGA combines the SDSS 2.5-meter telescope at Apache Point Observatory (Gunn et al. 2006) with the SDSS-BOSS spectrograph (Dawson et al. 2013; Smee et al. 2013; see Drory et al. 2015 for more details). This spectrograph has a wavelength range of 3600 to 10 300 Å and has an average spectral resolution of $R \approx 1800$.

MaNGA has 5 different configurations of hexagonal-formatted fiber bundles, which vary in diameter from 12".5 (19 fibers) to 32".5 (127 fibers; see table 2 in Bundy et al. 2015) to collect the light of galaxies up to $1.5R_{\rm eff}$ for galaxies in the Primary Sample and up to $2.5R_{\rm eff}$ for galaxies in the Secondary Sample, with a 2-to-1 split. The MaNGA sample selection is solely based on the galaxies' absolute $i$-band magnitude and redshift, with the final sample being characterized by an approximately flat distribution in the $i$-band magnitude and galaxy mass (for more information see Yan et al. 2019, and the discussion in Paper II).

### 2.2 The IllustrisTNG simulation suite

Illustris (Genel et al. 2014; Vogelsberger et al. 2014; Sijacki et al. 2015) is a suite of large-scale cosmological hydrodynamic simulations of galaxy formation and evolution. This first project forms the basis for IllustrisTNG (Marinacci et al. 2018; Naiman et al. 2018; Nelson et al. 2018; Springel et al. 2018; Pillepich et al. 2018b, 2019; Nelson et al. 2019a, 2019b). In the latter, the scientific goals are broader and new physics is introduced. Indeed, IllustrisTNG includes larger cosmological volumes (up to 300 Mpc instead of 100 Mpc), and higher-resolution runs with a mass resolution for baryonic matter up to $8.5 \times 10^4 \, M_\odot$ instead of $1.6 \times 10^6 \, M_\odot$. IllustrisTNG further includes physics of magnetic fields and dual-mode black hole feedback (Weinberger et al. 2017; Pillepich et al. 2018a), which are both not included in Illustris.

Many fundamental physical processes acting on a wide range of spatial and temporal scales must be included in cosmological simulations to predict the formation and evolution of galaxies, in terms of their stellar and gas chemical composition, star formation history, morphology, interaction with the environment through inflows and outflows, etc. The spatial and mass resolution achieved by TNG50 is among the highest in cosmological hydrodynamic simulations. However, subgrid physics still plays a role because many astrophysical phenomena occur on scales below the resolution limits of the simulations, such as, for example, gas cooling, star formation, evolution and chemical enrichment, super massive black holes accretion and feedback, and magnetic fields (see the discussion in Pillepich et al. 2018a, and reference therein).

IllustrisTNG simulates three physical box sizes, with cubic volumes of approximately 50 (TNG50), 100 (TNG100), and 300 (TNG300) Mpc side lengths and different resolutions, all assuming Planck cosmology from Ade et al. (2016).[1] This is the cosmological framework also assumed in the analysis presented in this paper.

---

[1] i.e. ΛCDM cosmology background, with matter density parameter $\Omega_{\rm m} = 0.31$; dark energy density parameter $\Omega_\Lambda = 0.69$; Hubble constant $H_0 = 100 h \, {\rm km \, s^{-1} \, Mpc^{-1}}$, with $h = 0.68$; matter power spectrum amplitude of $\sigma_8 = 0.82$ and spectral index $n_{\rm s} = 0.97$.

Since the goal of this project is to obtain a simulated sample of galaxies closely resembling the MaNGA catalogues in terms of galaxy selection criteria and observational characteristics, TNG50 is most appropriate. TNG50 matches the spatial resolution of the MaNGA data cubes with a pixel size of 0.5 arcsec, i.e. a spatial sampling raging from $\approx 100$ pc at $z \approx 0.01$ to $\approx 1.5$ kpc at $z \approx 0.15$ (see Section 2.1).

In Paper I, we discuss in detail how we generate MaNGA data cubes from the TNG50 simulations (see Section 3.1). The construction of the iMaNGA galaxy catalogue, in particular reproducing the flat distribution in AB *i*-band magnitude and mass, is described in detail in Paper II (see Section 3.2).

### 2.3 MaStar: SDSS-based stellar population models

To model light emitted by the simulated galaxies, we adopt the Maraston et al. (2020) stellar population models, which are based on the $\sim 60\,000$ stellar spectra collected by the MaNGA stellar library MaStar (Yan et al. 2019; Abdurro'uf et al. 2021) and energetics and synthesis methods as in Maraston (2005) and Maraston & Strömbäck (2011).

With its $\sim 60\,000$ stellar spectra, MaStar is the largest stellar library ever assembled to date. The spectra were collected with the MaNGA instrument described in Section 2.1. Therefore, since the observational set-up is the same, these stellar spectra have the same wavelength range, spectral resolution and flux calibration as the MaNGA data cubes.

In this project, we adopt an extended version of the Maraston et al. (2020) models. This updated version covers a wider range of stellar ages, with the youngest populations considered being $\sim 3$ Myr. Below this value, we adopt MappingsIII star-forming region models (Groves et al. 2008). In the MaStar models, 42 age and 9 metallicity values are provided. The metallicity range goes from $-2.25$ to $0.35$ dex. The Maraston et al. (2020) models also include 8 different values for the low-mass IMF slope, ranging between 0.3 and 3.8, with the Salpeter (1955)'s slope being 2.35. For more details, see Maraston et al. (2020) and Hill et al. (2021). In this project, we adopt these models, assuming the Kroupa (2002) IMF, both in the construction of the mock MaNGA data cubes and when running FIREFLY to recover the stellar populations' properties.

Adopting the MaStar models to light up the simulated galaxies, the iMaNGA data cubes have the same spectral properties as MaNGA spectra. Furthermore, by also using these models in the FIREFLY full spectral fitting procedure, we ensure the exclusion of any bias that would be caused by the adoption of different spectral models. With this approach, we aim to minimize the differences between the MaNGA and iMaNGA catalogues, so that those remaining are intrinsic to TNG50. In Paper II, we demonstrate how with this approach, we recover the intrinsic information in TNG50 when running the full-spectral fitting analysis at the $1\sigma$ level for the entire catalogue.

### 2.4 FIREFLY MaNGA VAC

N22VAC present the FIREFLY MaNGA VAC (Value Added Catalogue).[2] In this MaNGA VAC, the FIREFLY algorithm is adopted to obtain the stellar population properties for all 10 010 galaxies observed by MaNGA (Section 2.1). The catalogue provides spatially resolved properties such as stellar age, stellar chemical composition, star formation rates, dust attenuation, and stellar and remnant masses. The VAC also provides some global properties, for example, central values, median values within a given aperture, and so on. Also, stellar age and metallicity gradients are made available. For the stellar population properties, both locally and globally defined, both mass-weighted and light-weighted quantities are provided.

The results are obtained by running the full spectral fitting code FIREFLY using both the MaStar (Maraston et al. 2020) and the MILES (Sánchez-Blázquez et al. 2006; Maraston & Strömbäck 2011) stellar population models. N22VAC discuss the differences between the results obtained by FIREFLY employing these different stellar population models, and also a comparison with other analyses of the MaNGA catalogue, such as the results obtained employing PIPE3D (Sánchez et al. 2016; Sánchez et al. 2022). In this work, we use the FIREFLY MaNGA VAC in its MaStar version when comparing iMaNGA with MaNGA.

## 3 METHODOLOGY

The methodology of this work is introduced in our two previous papers of this series, which we will summarize in Sections 3.1 and 3.2. In this paper, we expand the catalogue presented in Paper II, adding TNG50 galaxies that fall into the MaNGA Secondary sample selection boundaries, adding 500 objects to the iMaNGA catalogue as described in Section 3.3. Finally, we introduce our methodology to determine T-morphology and inclination for the iMaNGA galaxies in Section 3.4 and Section 3.4.2, respectively.

### 3.1 Paper I: mock MaNGA galaxies

Here, we briefly summarize our methodology to generate and analyse mock MaNGA galaxies from the TNG50 simulations. This process is adopted for all the galaxies in our current sample.

When selecting a simulated TNG50 galaxy for post-processing, the light is modelled as discussed in Section 2.3. Emulating real MaNGA observations, the light is collected by a synthetic IFU instrument, with a pixel size of 0.5 arcsec. At this stage, we adopt a Field of View (FoV) of 150 arcsec per side. In this way, we can collect all the light emitted by the simulated galaxies. The kinematics are incorporated in the synthetic spectra, based on the stellar and gas kinematics provided by TNG50. Section 3.2 of Paper I contains a detailed description of this set-up. The IMASTAR code developed to perform these steps is publicly available.[3] To have a random viewing angle over the entire sample, we consider the *z*-axis of the cosmological volume as our line-of-sight (LoS), since the galaxies are randomly distributed within the simulation. The (random) inclination of galaxies in the iMaNGA catalogue is discussed in detail in Section 3.4.2.

To include the effects of dust in the synthetic data cubes, we run radiative transfer simulations with SKIRT (Baes et al. 2011; Baes & Camps 2015). These calculations are carried out twice at low spectral resolution for each galaxy, with and without dust included, in order to reconstruct the attenuation curve in each spaxel as the ratio between these two output data cubes as presented in section 3.2.2 of Paper I. With this approach, we avoid including Poisson noise in the spectra and significantly reduce computing time.

From the synthetic data cubes, we generate SDSS-like images (see section 3.3 in Paper I). The Sérsic 2D fitting code STATMORPH (Rodriguez-Gomez et al. 2019) is then used to derive galaxy morphology and the effective radius $R_{eff}$, which is essential for the implementation of proper MaNGA FoVs into our simulated data cubes.

---

[2] https://www.sdss4.org/dr16/manga/manga-data/manga-FIREFLY-value-added-catalog/

[3] https://github.com/lonanni/iMaNGA.

In MaNGA, 5 different hexagonal-fiber-bundle configurations are used to collect light within 1.5 $R_{eff}$ (in the MaNGA Primary sample) or within 2.5 $R_{eff}$ (in the MaNGA Secondary sample). These FoVs are adopted in iMaNGA to observe a galaxy with a given $R_{eff}$. We further consider the MaNGA effective PSF and mimic the typical noise in MaNGA observations, which is spatially and wavelength-dependent. These steps are discussed in detail in section 3.4 of Paper I. Thanks to this approach, the iMaNGA data cubes have the same characteristics as MaNGA data cubes in terms of spatial sampling, spatial resolution, spectral resolution, flux calibration, and noise.

Once we have the mock MaNGA data cubes, we follow the steps of the MaNGA DAP (Westfall et al. 2019) to analyse MaNGA galaxies. We employ the Voronoi algorithm by Cappellari & Copin (2003) with target $S/N_g > 10$, and then run the full-spectral-fitting code PPXF (Cappellari 2017) over the spectra in order to obtain the stellar kinematics, the gas kinematics, as well as the gas emission lines.

The full spectral fitting code FIREFLY (Wilkinson et al. 2017) is then used to derive the stellar population properties age and metallicity, as well as stellar mass, reddening and SFH. This is equivalent to what was done for observed MaNGA galaxies (see for example Goddard et al. 2016, 2017; Lian et al. 2018; Oyarzún et al. 2019, N21 and N22VAC).

### 3.2 Paper II: the iMaNGA sample

Here, we summarize the second Paper of this series. The focus of this paper is on the construction of the iMaNGA sample.

To build a MaNGA-Primary sample of galaxies in TNG50, we selected all galaxies in the magnitude and redshift range of the MaNGA survey. This initial sample includes all the galaxies in TNG50 within the MaNGA redshift range, i.e. $0.01 \le z \le 0.15$ (see Section 2.1). We rejected objects with fewer than 10 000 stellar particles, in order to ensure that galaxies are sufficiently resolved (see for example Schulz et al. 2020). As discussed in Paper II, there are 48 248 galaxies in TNG50 which satisfy these selection criteria.

The MaNGA-Primary sample is characterized by a smooth distribution in cosmological sampling. This is not the case in this initial sample, since TNG50 provides galaxies in redshift snapshots, hence with a discrete redshift distribution. Therefore, we randomly assign a redshift $z_{random}$ to each galaxy, $z_{random}$ being between the redshift of the galaxy's snapshot and the redshift of the next lower redshift snapshot. In this way, the redshift distribution is continuous. At this stage, we apply the MaNGA-Primary sample selection boundaries (see fig. 3 in Paper II), obtaining the 'parent sample'. To obtain a flat distribution in $i$-band AB magnitude, we randomly extract galaxies from this sample with higher probability when characterized by least probable magnitude values. This step leads to a final sample of ∼1000 galaxies. We refer the reader to the discussion in section 4 in Paper II for more detail.

### 3.3 Extending the iMaNGA sample

Next, we extend the iMaNGA sample selection to mimic the MaNGA Secondary Sample. As for the Primary MaNGA sample, for the latter, the selection criteria are uniquely based on galaxies' $i$-band AB magnitude and redshift, but selecting galaxies at higher redshift in order to increase the effective field of view of the IFU instrument.

Imposing the MaNGA Secondary sample selection criteria in TNG50 yields 889 galaxies in the 'parent sample'. From this 'parent sample', we generate the final sample with a flat $i$-band magnitude distribution following the procedure as described in Paper II. This final selection yields ∼500 galaxies, which exactly matches the ratio of 2:1 between the MaNGA Primary and Secondary Samples.

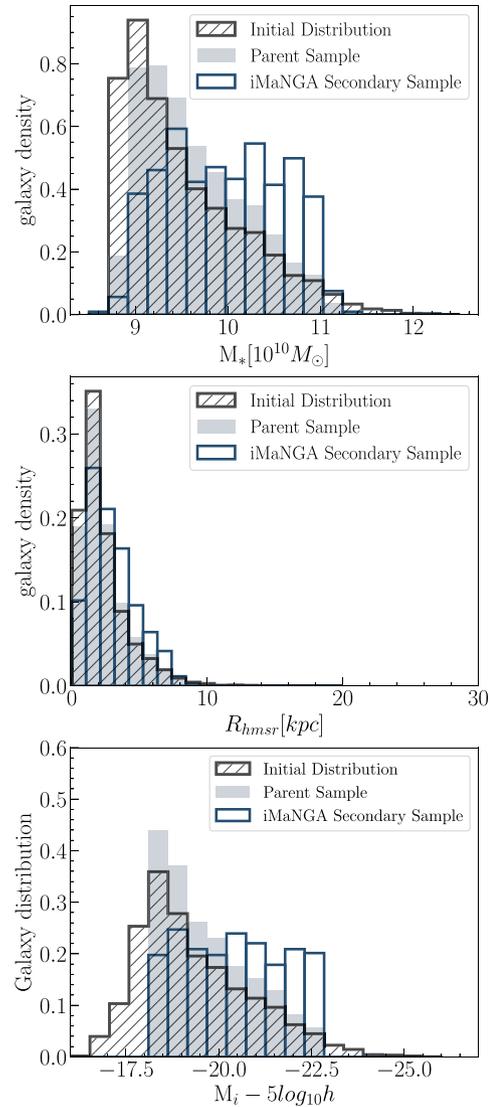

**Figure 1.** The distribution of the TNG50 galaxies from the initial sample (hatched black histograms), the parent sample (grey solid histograms), and the iMaNGA Secondary sample (blue empty histograms) in stellar mass (upper panel), half-mass-stellar radius (central panel) and $i$-band AB magnitude (bottom panel).

In Fig. 1, we present the distribution of the galaxies in the 'initial sample', 'parent sample', and iMaNGA Secondary sample, in terms of stellar mass (upper panel), half-mass-stellar radius (i.e. a proxy for the galaxy size; central panel), and the $i$-band magnitude (lower panel), mimicking figs 5–6 in Paper II for the iMaNGA Primary sample. It can be appreciated how approximately flat distributions in $i$-band magnitude and stellar mass are achieved.

Fig. 2 shows the distributions of galaxies in the MaNGA and iMaNGA secondary samples, in terms of the ratio between the FoV used to observe the galaxies and the effective radius, in the various total stellar mass bins. The overall distributions closely resemble each other (see grey histograms and quantiles). Some interesting differences appear in the skews. In particular, at the lower stellar mass bins ($\log_{10} M_*/M_\odot < 9.50$), there is a population of iMaNGA galaxies with a ratio $>10$, which does not exist in MaNGA. Interestingly, the MaNGA sample contains very diffuse objects in the highest mass bin ($log_{10} M_*/M_\odot > 11.0$) that are absent in the iMaNGA sample. These results are in line with what found in Paper II (cf. their fig. 10) about angular sizes in MaNGA and iMaNGA Primary Samples.

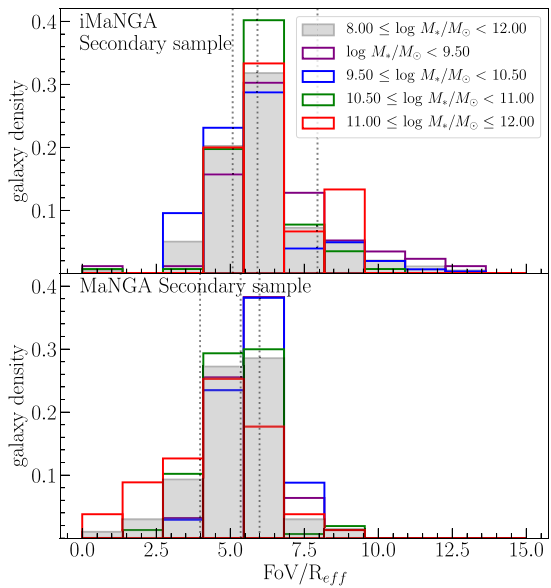

**Figure 2.** Distributions of FoV – effective radius ratio for iMaNGA (upper panel) and MaNGA (bottom panel) secondary sample galaxies, for the total sample (in grey) and in different total stellar mass bins (see the legend). Information for the MaNGA secondary sample galaxies is retrieved from DRPALL MaNGA data (Law et al. 2016). The 0.16, 0.50, 0.84 quartiles are reported (vertical dotted lines).

We post-processed and analysed all the TNG50 galaxies in the iMaNGA Secondary sample, following the method presented in Paper I, and as done in Paper II for the iMaNGA Primary sample. Hereafter, with 'iMaNGA sample', we will refer to the combination of the iMaNGA Primary sample and Secondary sample, characterized by 1511 galaxies.

All the unique mock MaNGA synthetic data cubes generated for this project are now publicly available on the IllustrisTNG website.[4]

### 3.4 Galaxy morphology & inclination

In this section, we discuss how we analyse and define the galaxy morphology, the stellar surface mass density, and galaxy inclination in the iMaNGA sample, following the method in N21 for the MaNGA catalogue.

#### 3.4.1 Galaxy morphology

We use STATMORPH (see Section 3.1) to derive galaxy morphologies and to calculate Sérsic indices and Petrosian radii (Sérsic 1963, 1968; Petrosian 1976). To determine the inclination as described in the following section, ellipticity and T-morphology are required. To this end, we adopt the Petrosian 'radius' ellipticity. In N21, T-morphology is based on Domínguez Sánchez et al. (2022). For our iMaNGA sample, we visually inspected all galaxies, dividing them into 4 categories: ETG, S0, LTG, and irregulars/merging (see also fig. 8 in Paper II for examples of these different morphologies in the iMaNGA sample). These correspond to T values: −3, 0, 3, 10, respectively. See Appendix B for a comparison between the T-values and the Sérsi index.

[4]https://www.tng-project.org/data/docs/specifications/.

#### 3.4.2 Galaxy inclination

As briefly discussed in Section 3.1, the simulated galaxies are observed by the mock IFU instrument with an LoS fixed to the z-axis of the cosmological volume. Therefore, observations of galaxies in the iMaNGA sample are characterized by random viewing angles. Here, we discuss how the inclination of the galaxies is computed for the iMaNGA sample.

We calculate the inclination between the assumed LoS and the galaxy-spin axis provided by TNG50. The total spin of the galaxies is computed in TNG50 as the mass-weighted sum of the relative coordinate times the relative velocity of all member particles/cells. This definition of the inclination is, therefore, kinematics-dependent and also 'theoretical', being based on the intrinsic information in TNG50, as opposed to observational information. To compute this 'theoretical' inclination, we measure the inclination between the viewing angle, that is the z-axis of the cosmological volume, and the spin of each galaxy, i.e. $J = [J_x, J_y, J_z]$. We will refer to this inclination as 'kinematic-dependent inclination' or 'theoretical inclination'.

In N21, instead, the inclination is computed from the galaxy's morphology, assuming

$$cos(i) = \sqrt{\frac{(q^2 - q_0^2)}{(1 - q_0^2)}}, \quad (1)$$

where $i$ is the inclination, $q_0$ is the intrinsic thickness of the galaxy, and $q$ is the observed axial ratio of the projected spheroid (Hubble 1926). For MaNGA galaxies, $q$ is obtained from the elliptical Petrosian analysis (Wake et al. 2017a). Assuming that the intrinsic axial ratios only vary with morphology and that galaxies seen face-on are perfectly circular, N21 find the relation

$$-\log q_0 = 0.316 + 0.049T, \quad (2)$$

where T is the T-value for the T-morphology (see Domínguez Sánchez et al. 2022). For more information, see N21 and the references therein.

To mimic this approach in the present work, we assume equations (1–2), the T-morphology of the iMaNGA galaxies, and the Petrosian axial ratio as defined above. We will refer to this inclination as 'morphology-dependant inclination'.

Fig. 3 shows galaxy inclinations for the following three morphological types: ellipticals (E), lenticular (SO), and spirals (LTG). The MaNGA and iMaNGA samples are shown as empty and filled histograms, respectively. For the iMaNGA sample, we show both the kinematics-dependent (hatched histograms) and the morphology-dependent (solid histograms) inclinations. It can be noticed that the theoretical inclinations and the morphology-dependent inclinations do show an overall agreement, but differences appear, in particular at higher inclinations for early-type galaxies.

The differences between 'theoretical' and 'morphology-dependant' inclinations are not surprising. The latter, which is based on axis ratios, can only be approximate, as galaxies are not infinitely thin or perfectly round (see Ryden 2004, and the discussion in N21). Therefore, N21 account for an intrinsic thickness; we refer the reader to the discussion therein.

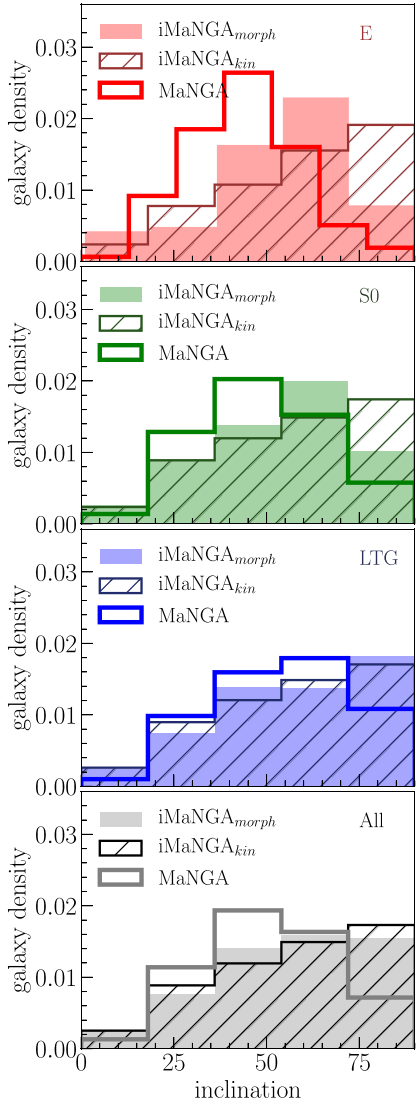

**Figure 3.** The distribution of galaxies' inclination in the iMaNGA and MaNGA galaxy catalogue, divided into 3 morphological types (i.e. E, S0, and LTG, in the first three panels) and for the entire sample (last panel). The inclinations for the MaNGA sample are computed as in N21 (empty histograms). For the iMaNGA sample, the inclinations are computed following the same method as in N21, i.e. based on the T-morphology (solid histograms), and from the kinematics in TNG50 (hatched histograms), as explained within this section.

The distributions of the morphological inclinations in MaNGA and iMaNGA agree well overall, although the density of edge-on galaxies is higher in iMaNGA compared to the MaNGA sample. This discrepancy is likely caused by the fact that we have not applied any cuts based on inclination, galaxy size, or dust in iMaNGA, while such observational selection effects will play a role in MaNGA. For instance, edge-on galaxies are more rarely observed in MaNGA because of obscuration effects (see the discussion in Wake et al. 2017b; Abril-Melgarejo et al. 2021, and N21). When looking at the inclination distribution for the different morphological types, we find a discrepancy for S0 to E, with only the LTG distributions agreeing well. This discrepancy can be explained by two effects. First, in order to determine the morphological inclinations, we apply equation (2) from N21, which is based on the T-values for the MaNGA sample.

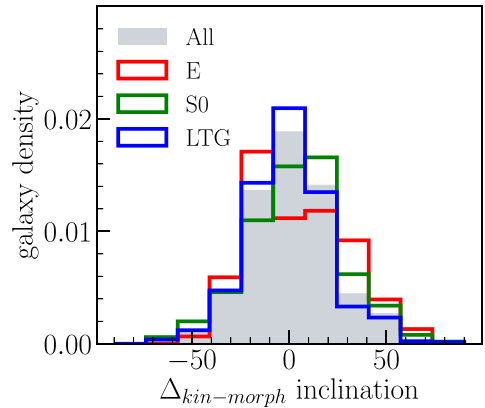

**Figure 4.** The distribution of the differences between the galaxy inclinations considering the morphology-dependent and the kinematics-dependent determinations of galaxies' inclinations.

As discussed above, in this analysis, we consider – for simplicity – only 4 possible T-values, namely −3, 0, 3, and 10 for E, S0, LTG and irregulars, respectively (see the discussion above and Appendix B). In N21, instead, the T-value follows a continuous distribution with, in particular, the majority of elliptical galaxies having T-value > −3 (see fig. 2 in N21). Hence, with this simplification, we are assuming that elliptical objects in the sample have all the same intrinsic thickness. In addition, there might be intrinsic differences between the S0 and E galaxies in TNG50 and MaNGA. This intrinsic difference is discussed in Paper II as well as later in this paper.

Fig. 4 reports the difference between the theoretical inclinations and the morphology-dependent inclinations for the iMaNGA sample. Overall, the inclinations based on kinematics appear slightly higher than those based on morphology. To emulate the observationally determined inclination in N21, we will adopt the morphology-dependent inclination for both iMaNGA and MaNGA in the analysis.

### 3.4.3 Stellar surface mass density

The stellar surface mass density $\Sigma_*$ is calculated by dividing the total stellar mass by the surface area. In the FIREFLY VAC (see N22VAC), the stellar masses are provided by FIREFLY in each Voronoi tassel. In N21, the area of the Voronoi tassels is corrected for projection effects. Therefore, the area is $A = A_{\rm obs} sec(i)$, where $A_{\rm obs}$ is the observed area. Consequently, the surface mass density corrected for projection effects is $\Sigma_* = \Sigma_{*, {\rm obs}} cos(i)$. This is the quantity that we will show throughout the paper for both the MaNGA and the iMaNGA samples.

Fig. 5 illustrates the distribution of stellar surface mass density (left panels) and stellar mass (right panels), in both the iMaNGA and MaNGA samples, divided into morphological types (E, S0, and LTG, in the first three rows) and for the entire sample (last row). It can be noticed how the distributions closely resemble each other, with the iMaNGA sample characterized by lower stellar masses in each morphological bin and therefore slightly lower stellar surface mass density. The lack of massive spheroidal galaxies has already been noticed and discussed in Paper II (see discussion and references therein). For both catalogues, to enhance the comparison, we consider the total stellar mass as obtained by running FIREFLY (see Appendix A for a discussion of other definitions for the total stellar mass in these samples).

The comparisons shown in Fig. 4 demonstrate the validity of the following analysis, based on the definition of the inclination and deprojected stellar surface mass density for the MaNGA and iMaNGA samples, since no particular bias is introduced in the overall iMaNGA sample. The smallest fraction of massive Es and S0s in iMaNGA seen in Fig. 5 is intrinsic to this simulation, as already discussed in Paper II.

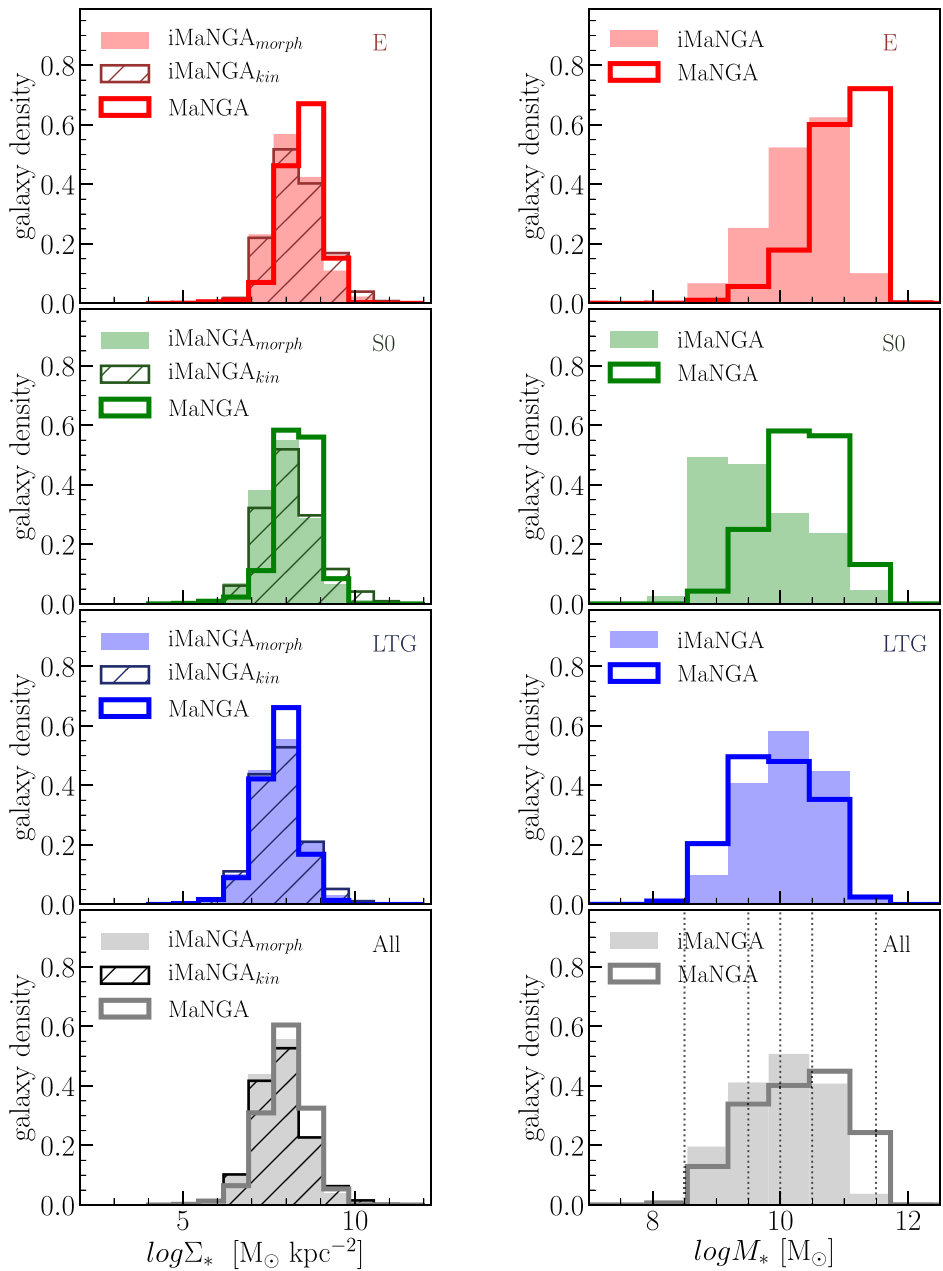

**Figure 5.** The distribution of galaxies' stellar surface mass density $\Sigma_*$ (left panels) and total stellar mass (right panels) in the iMaNGA and MaNGA catalogues. The distributions are divided into 3 morphological types (i.e. E, S0, and S, in the first three panels) and the entire sample (bottom panels). The stellar surface mass density $\Sigma_*$ and stellar mass, for both samples, are computed for each galaxy by running the full-spectral-fitting code FIREFLY on each Voronoi tassel. These data from the MaNGA sample are retrieved from the FIREFLY VAC, see N22VAC. In the bottom right panel, the vertical dotted lines represent the extremes of the bins in stellar mass that are used in the following analysis; see Table 1.

## 4 RESULTS

In this section, we present the results of the direct comparison between the iMaNGA and the MaNGA sample. We focus on the stellar population properties, age and metallicity, in the three-dimensional parameter space of stellar mass, stellar surface mass density, and galaxy radius. For the MaNGA sample, we use the data provided by the FIREFLY VAC in its MaStar run (see Section 2.4). The stellar population properties for the iMaNGA sample are calculated using the full spectral fitting code FIREFLY with the same settings as for the MaStar-based FIREFLY VAC (see Section 3.1) to ensure consistency. We show light-weighted stellar age and stellar metallicity, if not noted differently. Figures from N21 are replicated, however, it is also important to note that N21 report findings obtained using the MILES

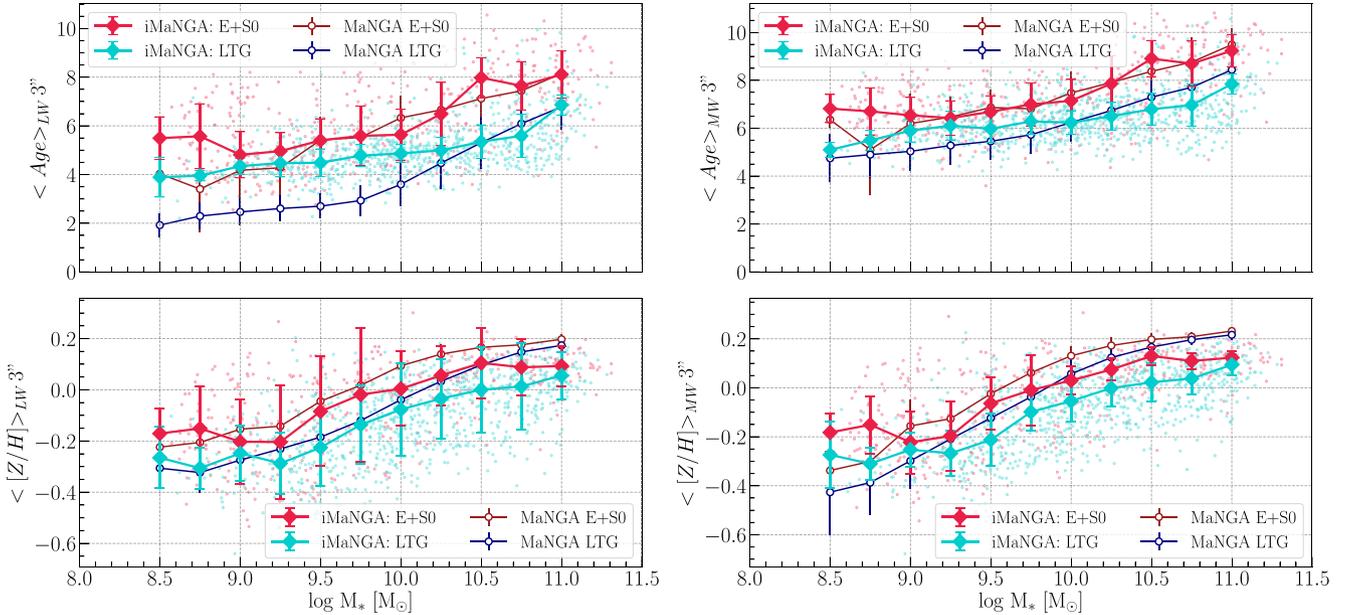

**Figure 6.** The global stellar mass–age (top panels) and –metallicity relation (bottom panels) separated by morphology, for the MaNGA and iMaNGA samples. The left panels show light-weighted quantities, and the right panels show mass-weighted ones. Age and metallicity are averaged within a central area of 3 arcsec diameter. The scatter points display each galaxy in the iMaNGA sample and the line plot shows the median age and metallicity across mass bins of 0.25 dex width, for the iMaNGA (solid diamonds) and MaNGA (empty circles) galaxies. The E+S0 galaxies are represented in red, while the LTGs are represented in blue. The error bars illustrate the standard deviation.

SSP models (Sánchez-Blázquez et al. 2006) in FIREFLY. Furthermore, there are general differences in the way data are analysed in this work compared to N21. The data analysis followed here is explained in full within the Sections. As in N21, we exclude galaxies that present signs of mergers, with an inclination angle >80 deg, with fewer than 30 Voronoi tassels, and reject Voronoi tassels with SNR < 8.

### 4.1 Stellar population scaling relations

Fig. 6 shows the global stellar mass–age and mass–metallicity relations in early-type galaxies (ETGs) and late-type galaxies (LTGs) in the iMaNGA and MaNGA sample, mimicking fig. 4 of N21. Here, we report both light-weighted (on the left) and mass-weighted (on the right) quantities. All galaxies in the iMaNGA sample are shown as small dots in the background. The median of stellar age and metallicity within a 3 arcsec diameter aperture is measured in bins of mass of 0.25 log $M_*$ dex. The error bars illustrate the standard deviation in each mass bin, to better illustrate the scatter of the distributions of galaxies in the samples.

The observed positive mass–age relations of the MaNGA sample for both ETGs and LTGs are well matched by the iMaNGA sample, with most massive galaxies being older. The difference between galaxy types is generally reproduced in iMaNGA with late-type galaxies having younger stellar population ages. However, the difference is predicted to be slightly smaller by the simulations. The only significant discrepancy can be seen for light-weighted ages of lower-mass late-type galaxies with the simulated galaxies being up to 2 Gyrs older than the corresponding observed ones.

The observed global stellar mass–metallicity relations, referred to as MZR in literature, are generally well reproduced in iMaNGA with galaxy metallicity increasing with galaxy mass for both ETGs and LTGs. The simulations further reproduce well the fact that metallicities are slightly lower in LTGs than in ETGs, even though the observed larger discrepancy at low masses is less pronounced in iMaNGA. We notice that the simulations slightly underestimate the metallicities of the most massive galaxies by about 0.1 dex for ETGs and 0.2 dex for LTGs.

Both nebular and stellar MZR have been investigated many times in the literature for observed galaxies (for example, Tremonti et al. 2004; Gallazzi et al. 2005). In recent years, the global MZR has also been studied in simulations. For example, Torrey et al. (2019) investigated the gas-phase MZR in TNG100, finding good agreement with the observation looking at galaxies with a total stellar mass >$10^9$ $M_\odot$ and redshift between 0 and 2. Similarly, De Rossi et al. (2017) quantified the global gas and stellar MZR in the EAGLE simulations. These relations agree well with the observational findings up to redshift 3. Furthermore, Cook et al. (2016) produced the stellar MZR for simulated galaxies in Illustris, in particular, focussing on ETGs with stellar mass >$10^{10}$ $M_\odot$ finding good agreement with observations with the shape, but presenting overall lower metallicity, as found in this analysis. In these works, data in the simulations are utilised without forward modelling of the simulated galaxies to bring them into the observational plane. These trends are also investigated in Nelson et al. (2018), where galaxies in TNG100 and TNG300 at redshift 0.1 are selected and compared with observations at the same redshift (Gallazzi et al. 2005, in particular). For the stellar MZR in particular, SDSS-like spectra were generated for the comparison, finding an offset in the metallicity trends compared to observations of up to 0.5 dex at lower stellar mass. Therefore, while in this work, excellent agreement for the galaxy colours was found between observations and simulations, the comparison between stellar ages and metallicity showed some tensions, concluding that this more direct comparison is somehow more difficult because of the necessity to bring the simulated galaxies into the observational plane in a rigorous manner.

**Table 1.** The number of galaxies and the mean stellar mass in each bin in morphology and total stellar mass for both the MaNGA and iMaNGA galaxy catalogue. These bins constitute the mass-morphology plane adopted throughout the paper.

|  |  | $8.50 \leq \log M_*/M_\odot < 9.50$ | $9.50 \leq \log M_*/M_\odot < 10.0$ | $10.0 \leq \log M_*/M_\odot < 10.50$ | $10.50 \leq \log M_*/M_\odot \leq 11.50$ |
|---|---|---|---|---|---|
| E | iMaNGA | $N_{\rm gal} = 7$; $<M_*> = 9.14$ | $N_{\rm gal} = 17$; $<M_*> = 9.76$ | $N_{\rm gal} = 24$; $<M_*> = 10.29$ | $N_{\rm gal} = 26$; $<M_*> = 10.78$ |
|  | MaNGA | $N_{\rm gal} = 44$; $<M_*> = 9.24$ | $N_{\rm gal} = 111$; $<M_*> = 9.8$ | $N_{\rm gal} = 253$; $<M_*> = 10.28$ | $N_{\rm gal} = 1947$; $<M_*> = 10.8$ |
| S0 | iMaNGA | $N_{\rm gal} = 122$; $<M_*> = 9.06$ | $N_{\rm gal} = 40$; $<M_*> = 9.75$ | $N_{\rm gal} = 36$; $<M_*> = 10.26$ | $N_{\rm gal} = 39$; $<M_*> = 10.89$ |
|  | MaNGA | $N_{\rm gal} = 64$; $<M_*> = 9.22$ | $N_{\rm gal} = 167$; $<M_*> = 9.77$ | $N_{\rm gal} = 279$; $<M_*> = 10.27$ | $N_{\rm gal} = 363$; $<M_*> = 10.86$ |
| LTG | iMaNGA | $N_{\rm gal} = 122$; $<M_*> = 9.22$ | $N_{\rm gal} = 172$; $<M_*> = 9.78$ | $N_{\rm gal} = 180$; $<M_*> = 10.24$ | $N_{\rm gal} = 169$; $<M_*> = 10.81$ |
|  | MaNGA | $N_{\rm gal} = 1293$; $<M_*> = 9.18$ | $N_{\rm gal} = 1128$; $<M_*> = 9.74$ | $N_{\rm gal} = 1171$; $<M_*> = 10.24$ | $N_{\rm gal} = 1074$; $<M_*> = 10.78$ |

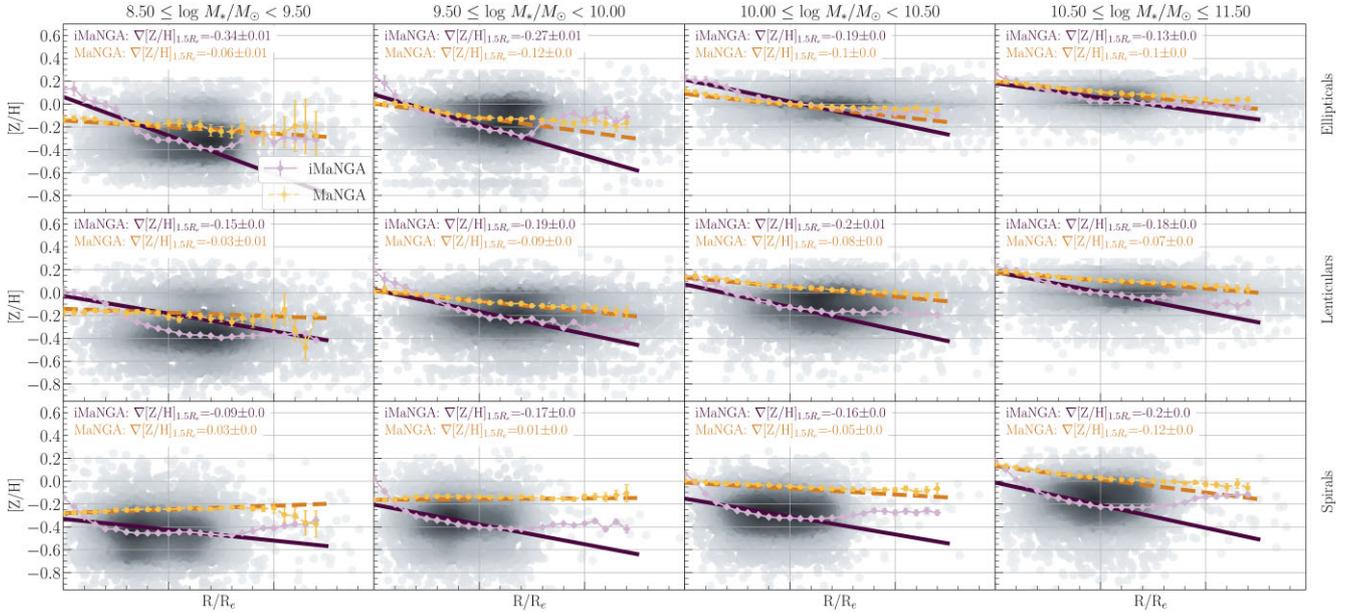

**Figure 7.** Radial metallicity profiles, dividing the iMaNGA and MaNGA sample in stellar mass (columns) and morphology (rows) bins; see Table 1. [Z/H] is recovered with FIREFLY for both samples in the same manner. In each panel, we show the median [Z/H] in 0.1 $R_{\rm eff}$ width bins for iMaNGA (pink diamonds) and MaNGA (orange circles), considering all spaxels up to 2.5 $R_{\rm eff}$. The error bars represent the standard error on the median, see equation (3). Linear regressions are presented up to 2.5 $R_{\rm reff}$, computed on data up to 1.5 $R_{\rm reff}$ (solid violet line for iMaNGA, orange dashed lines for MaNGA). Gradients are reported in the top-left corner of each panel for both catalogues. In the background of each panel, we show the distribution of the galaxies in the iMaNGA sample, calculated with a Gaussian kernel density estimator.

### 4.2 Radial profiles

In this section, we investigate the radial profiles of stellar metallicity, age, and stellar surface mass density. Figs 7 and 8 show the radial profiles of age and metallicity in different mass-morphology bins, respectively. Fig. 9 shows the equivalent for the stellar surface mass density. The shadows in each panel are the iMaNGA galaxy density distribution, calculated with a Gaussian kernel density estimator. Median values in bins of 0.1 $R_{\rm eff}$ are shown for both iMaNGA and MaNGA (see labels). The error bars represent the standard error on the median, defined as:

$$\sigma_{\rm err} = \frac{\pi}{2} \frac{\sigma}{\sqrt{N}}, \quad (3)$$

where $\sigma$ is the standard deviation in the bin and $N$ is the number of data that populate the bin. Only light-weighted quantities are used. The gradients reported in the upper left corner of each panel are calculated as discussed in section 4.1.2 in Paper I. Gradients are computed considering all data up to $1.5R_{\rm eff}$ and the linear regression lines are reported for both catalogues up to $2.5R_{\rm eff}$.

In Table 1, we list the characteristics of the mass-morphology bins used throughout the analysis presented in this paper. In particular, we report the number of galaxies and their average stellar mass for each mass-morphology bin.

#### 4.2.1 Stellar metallicity

The radial profiles in stellar metallicity are shown in Fig. 7. We can see that, overall, the iMaNGA gradients are steeper than the corresponding MaNGA ones. As also noted in Fig. 6, the MaNGA sample is also populated by more metal-rich stellar populations overall. It is striking that overall the metallicity distribution between iMaNGA and MaNGA agree reasonably well for elliptical galaxies, except for the lowest-mass bin. The picture is different for lenticular and spiral galaxies: while the central values are in reasonable agreement (see also Fig. 6), simulations predict significantly steeper radial metallicity gradients in most mass bins. This implies that metallicities around the effective radius and beyond are considerably underestimated in the simulations by up to 0.2 dex.

For both catalogues, we can observe that along the columns, i.e. at fixed stellar mass, the metallicity increases going from spirals to ellipticals; looking along the rows, i.e. at fixed morphology, the metallicity increases going towards higher stellar mass.

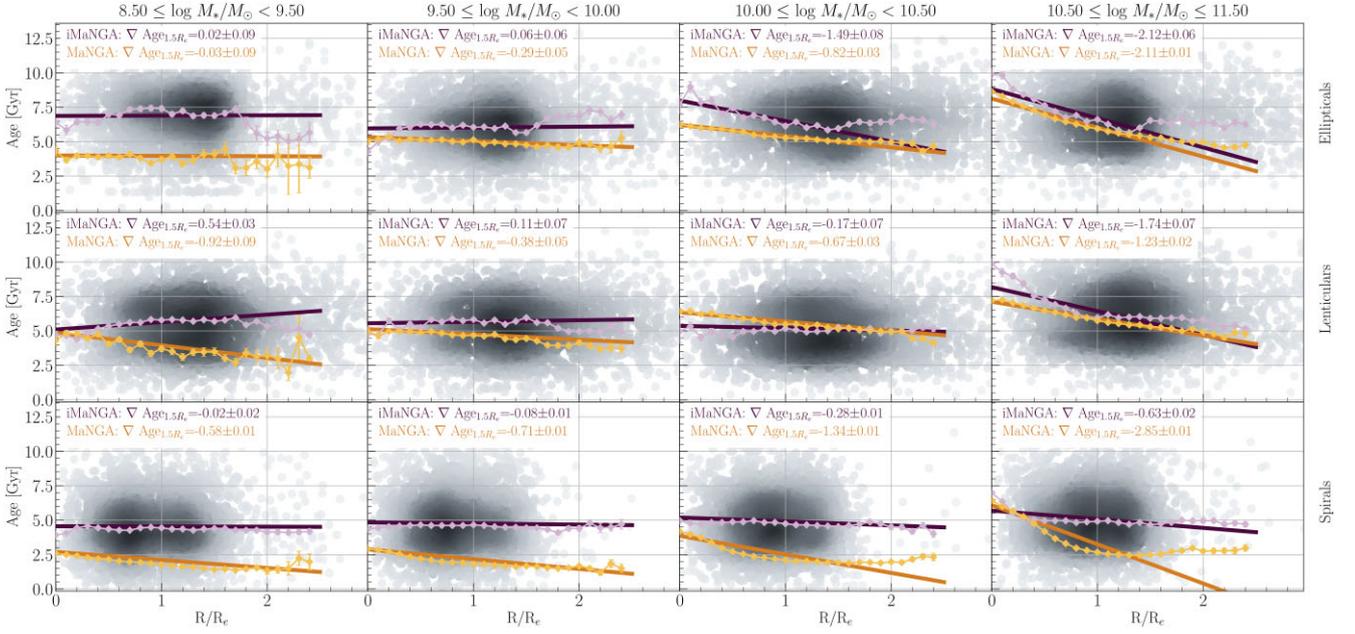

**Figure 8.** As Fig. 7, this time considering the stellar age.

An important result found with MaNGA data is that the generally negative metallicity gradients significantly steepen with increasing galaxy mass with flat and potentially positive gradients in the lowest-mass galaxies at around $\log M_* \sim 9 M_\odot$ (see also Goddard et al. 2016, 2017, and N21). In other words, in MaNGA, there is a break in the negative trend of the [Z/H] with the stellar mass for low-mass galaxies. In particular, as noted in Goddard et al. (2017), the dependence on the mass is stronger for spiral galaxies.

Looking at elliptical galaxies in iMaNGA, as also observed in Cook et al. (2016), the trend with the stellar mass is absent: indeed, the bin at the lowest stellar mass is dominated by the strongest negative radial trends. Looking instead at lenticulars and spirals, what is seen in the literature is observed: moving toward higher stellar masses, the gradients are steeper, and this trend is stronger for spiral galaxies.

We will further discuss the relation between the metallicity gradients and the galaxy stellar mass in Section 4.4.

*4.2.2 Stellar age*

Fig. 8 shows the age radial profiles. As for Fig. 6, iMaNGA is characterized by older stellar populations overall compared to MaNGA. While age gradients in MaNGA are only negative or flat, iMaNGA presents negative or flat gradients, with the exception of a positive gradient for lenticular galaxies with stellar mass between 8.5 and 9.5 $M_\odot$. Cook et al. (2016) also observe not only negative gradients for the stellar age in Illustris for ETGs.

The comparison for age provides quite a different picture than the comparison for metallicity. While simulations and observations agree quite well for metallicity in elliptical galaxies, as discussed above, there is a stark discrepancy for age. Indeed, except in the highest mass bin, the simulations predict higher light-weighted stellar ages at all radii in elliptical galaxies. The discrepancy is as high as 4 Gyr for the lowest mass ellipticals. However, the age gradients are consistent: both samples present flat age gradients for low- and intermediate-mass galaxies, and negative gradients at $\geq 10 M_\odot$.

A similar pattern can be seen for lenticular galaxies, although slightly less pronounced. The largest discrepancy between simulation and observation can again be seen in spiral galaxies. iMaNGA galaxies exhibit ages systematically higher by about 2.5 Gyr for all galaxy masses, at all radii. The discrepancy tends to worsen to larger radii, with the simulations predicting slightly flatter age gradients than what is observed in MaNGA. This result mirrors the metallicity profiles in spiral galaxies, where we see steeper gradients in the simulations compared to observations as discussed above.

The gradients here discussed consider light-weighted quantities. Mass-weighted results are presented in Section C. The overall results remain unchanged for mass-weighted quantities. In Section C, we also discuss our ability to recover the 'intrinsic' TNG50 information, and the 'intrinsic' gradients, in the mass-morphology plane, demonstrating in this way how the 'recovered' gradients are compatible with the 'intrinsic' ones, i.e. the differences between iMaNGA and MaNGA are not generated by the forward-modelling approach adopted but are intrinsic to TNG50.

*4.2.3 Stellar surface mass density*

Fig. 9 shows the radial profile of the stellar surface mass density, $\Sigma_*$. $\Sigma_*$ is computed for both samples considering the stellar mass as outputted by FIREFLY and corrected for the projection effect considering the morphology-dependent determination of the inclination of the galaxies (see Section 3.4.2). The gradients for the stellar surface mass density are in reasonable agreement for spiral galaxies, both in terms of shape and normalization. Instead, iMaNGA predicts steeper $\Sigma_*$ gradients in elliptical and lenticular galaxies. The mass-dependence observed in MaNGA for these galaxy types, a steepening of the negative $\Sigma_*$ gradient with increasing mass, is not entirely recovered by the simulations.

This result is partially consistent with the findings of Cannarozzo et al. (2022) that compare ETGs drawn from TNG100 with MaNGA observations and find agreement between the $\Sigma_*$ profiles of the simulated and observational data.

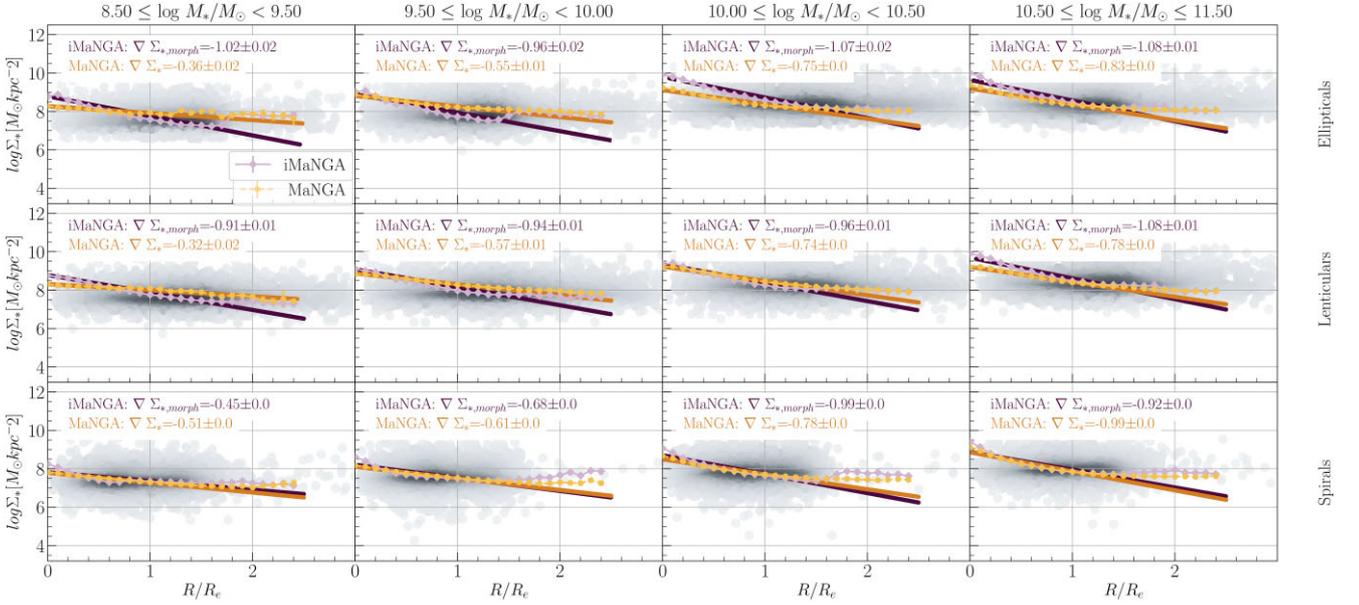

**Figure 9.** $\Sigma_*$-radius trends in the MaNGA and iMaNGA sample, in the morphology-stellar mass plane (see Table 1). $\Sigma_*$ is computed from the stellar mass recovered with FIREFLY and considering the inclination from the T-morphology for both samples, using the FIREFLY VAC data set (see N22VAC) for the MaNGA galaxies. In each panel, we show the median $\Sigma_*$ in 0.1 $R_{\rm eff}$ width bins, for iMaNGA (pink diamonds) and MaNGA (orange circles) galaxies, up to 2.5 $R_{\rm eff}$. The error bars represent the standard error on the median, see equation (3). The linear regressions to the data up to 1.5 $R_{\rm reff}$ are presented (solid violet line for iMaNGA, orange dotted lines for MaNGA). The gradients are reported on the top left corner of each panel for both iMaNGA and the MaNGA galaxies. In the background of each panel, we show the distribution of the galaxies in the iMaNGA sample, calculated with a Gaussian kernel density estimator.

As noted in N21, since both the stellar metallicity and the stellar surface mass density present negative radial trends, a relation within these two quantities must exist locally. We have found these two quantities to be characterized exclusively by negative trends in the iMaNGA catalogue. In MaNGA instead, flatter radial trends for the metallicity are found for low- and intermediate-mass galaxies. Following N21, we will therefore now investigate the interplay between these two quantities to shed light on the local drivers of stellar metallicity.

### 4.3 Dependence on surface mass density and radius

N21 investigate the spatially resolved stellar surface mass density–metallicity relation for MaNGA galaxies. Here, we explore this relation in the iMaNGA sample following the analysis steps of N21 and directly compare with MaNGA.

#### 4.3.1 Metallicity as a function of surface mass density

Fig. 10 shows the $\Sigma_*$-metallicity relation for both iMaNGA and MaNGA galaxies. All spaxels within $3R_{\rm eff}$ with SNR > 8 are included. The contours indicate the 20, 40, 60, 80 percentiles for the iMaNGA sample (dashed white lines) and for the MaNGA sample (solid orange lines). We further show the median stellar population metallicity in 0.1 dex bins in log $\Sigma_*$ for both samples, as well as the linear regression lines (see legend). The corresponding equations are in the upper left corner. All galaxies in both samples are considered, combining all morphological types and all stellar masses (see Table 1).

In both MaNGA and iMaNGA, we find a clear positive correlation between $\Sigma_*$ and [Z/H]. Hence, surface mass density is identified as a significant driver of stellar metallicity in MaNGA, and this is well reproduced in the simulations. The slopes of these relationships are comparable, the observed relation being slightly steeper. Most strikingly, however, there is a clear offset between these two relationships. The simulations systematically predict lower stellar metallicities by almost a factor 2 (0.25 dex) across all surface mass densities.

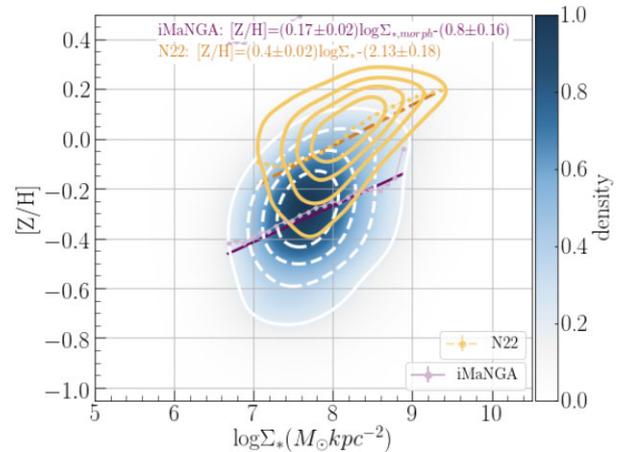

**Figure 10.** Local $\Sigma_*$–[Z/H] relation for MaNGA and iMaNGA galaxies, considering together all morphology and mass bins in Table 1. We present the density plot of all spaxels in the iMaNGA sample up to 3 $R_{\rm eff}$, smoothing the data with a Gaussian kernel. $\Sigma_*$ is corrected for the projection effect by assuming the 'morphology-dependent inclinations' for both catalogues. The contour lines enclose 20, 40, 60, and 80 per cent of the data for the MaNGA sample (orange lines) and the iMaNGA sample (white lines). The median of [Z/H] in 0.1 dex width bins in log $\Sigma_*$ is represented by violet diamonds for iMaNGA and with orange circles for MaNGA; the error bars (reported in the same colours) represent the standard error on the median. The linear regressions are represented (violet for the iMaNGA sample, and orange for the MaNGA galaxies). The gradients are reported in the top left corner of the panels.

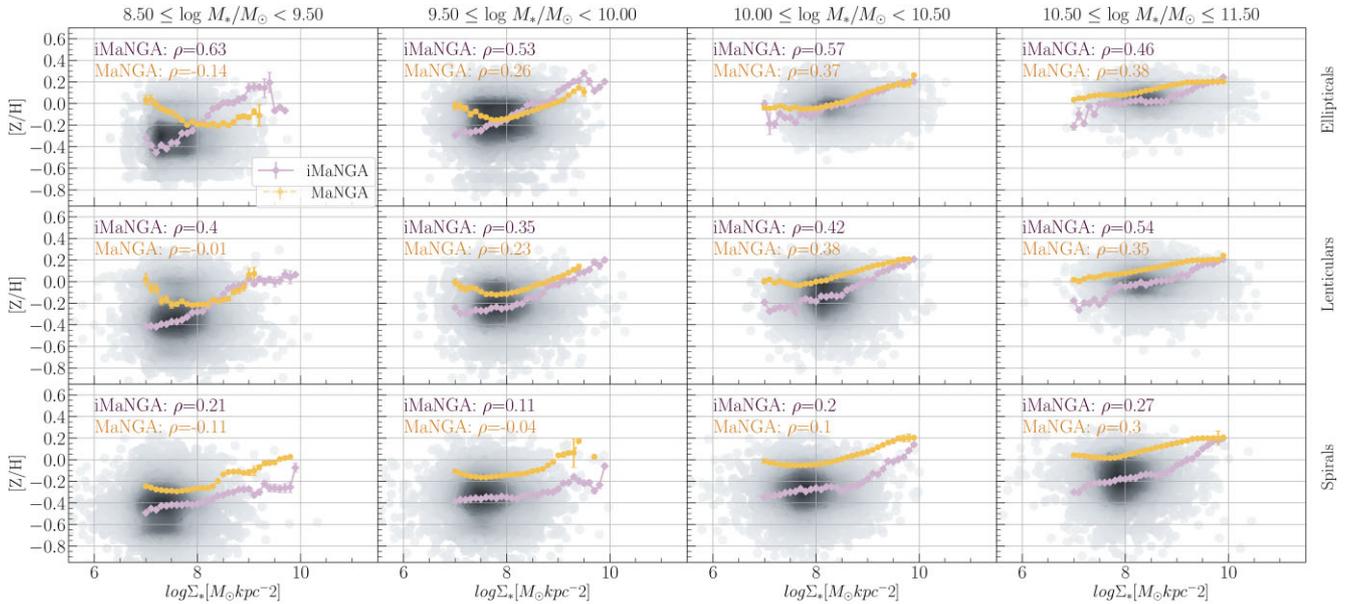

**Figure 11.** $\Sigma_*$–[Z/H] trends in the MaNGA and iMaNGA sample, in the morphology-stellar mass plane (see Table 1). $\Sigma_*$ is computed as discussed in Section 3.4.2, considering the correction for the 'morphological-dependent' inclination for the iMaNGA sample. In each panel, we show the median stellar metallicity in 0.1 log $\Sigma_*$ dex width bins, for iMaNGA (pink diamonds) and MaNGA (orange circles) galaxies, considering all the spaxels up to 3 $R_{\rm eff}$. In the background, we report the iMaNGA galaxy density distribution, calculated with the Gaussian kernel density estimator. The error bars represent the standard error on the median, see equation (3). In each panel, we report the Pearson coefficient in the upper-left corner for both the iMaNGA and MaNGA galaxies.

*4.3.2 Metallicity as a function of surface mass density in the morphology-mass plane*

In Fig. 11, we show the $\Sigma_*$–[Z/H] relation split by galaxy mass and morphology for both iMaNGA and MaNGA. The grid shows the global mass–morphology plane (columns for the stellar mass, rows for the galaxy morphology as described in Table 1). In particular, $\Sigma_*$ on the *x*-axis and the stellar metallicity [Z/H] on the *y*-axis are computed for both samples on the basis of the FIREFLY MaStar run (see N22VAC).

The positive correlation between surface mass density and metallicity is again apparent for both simulations and observations. As already shown in Fig. 10, we can further see that the metallicities in iMaNGA tend to be lower than in MaNGA across the full range of surface mass density. However, Fig. 11 provides us with more information on secondary dependencies on the mass and morphology of galaxies.

Most interestingly, the discrepancy between simulations and observations is strongest for late-type galaxies (up to ∼0.25 dex), a trend we have already noticed from the radial metallicity profiles presented in Fig. 7. MaNGA and iMaNGA instead agree well for elliptical galaxies, and lenticular galaxies seem to sit in between these two extremes.

At a further level of detail, we notice that metallicities are consistent at the highest surface mass densities for all galaxy morphologies. Furthermore, both samples present stronger positive relations between $\Sigma_*$ and [Z/H] at higher mass (see the Pearson correlation coefficient reported in the figure). However, the positive correlation tends to be stronger in the simulations than in the observations, especially in the lowest stellar mass bin and in spiral galaxies at any mass. There is no particular dependence on the galaxy mass, except that the $\Sigma_*$–[Z/H] relation of lower-mass galaxies appears to turn around in MaNGA. Indeed, in MaNGA, metallicity increases again toward the lowest $\Sigma_*$ values for galaxies with $M_* < 10^{10} {\rm M_\odot}$ and this effect is not displayed by iMaNGA. In other words, this break of the relation at the low stellar mass bins ($M_* < 10^{10} {\rm M_\odot}$) is absent in iMaNGA. This rise is difficult to interpret and may be driven by radial effects rather than surface mass density – see the discussion in N21 on additional radial-dependent drivers of metallicity in low-mass galaxies in MaNGA. The next section will shed more light on this question, where both parameters are discussed simultaneously.

*4.3.3 The radius–surface mass density plane*

From the analysis presented so far, we know that metallicity correlates with both surface mass density and galactocentric radius. However, the surface mass density also correlates with the galactocentric radius. Hence, the true drivers of local metallicity trends within galaxies can only be identified by analysing both parameters simultaneously.

Interestingly, N21 find in MaNGA that metallicity predominantly depends on the stellar surface mass density locally, with a strong positive correlation between $\Sigma_*$ and [Z/H] at any fixed radius. At fixed surface mass density, instead, no radial dependence is found in massive galaxies, while an interesting secondary dependence is detected in galaxies with stellar mass $\leq 10.80 {\rm M_\odot}$: metallicity *increases* with increasing radius. The implication of this result is that the negative correlation found previously between metallicity and radius is actually driven by the correlation between radius and surface mass density. In the following, we repeat this particular analysis with iMaNGA to test whether the TNG50 simulations reproduce this pattern observed with MaNGA.

Fig. 12 shows the radial $\Sigma_*$ profiles of the iMaNGA sample, colour-coded by median stellar metallicity. The figure mimics fig. 10 in N21. Again, we split by galaxy mass (columns) and morphology (rows). As in N21, the LOESS algorithm by (Cappellari et al. 2013) is adopted to better illustrate the underlying trends (see N21 for more details). For each mass-morphology bin, as in N21, we also report the minimum and maximum stellar metallicity [Z/H].

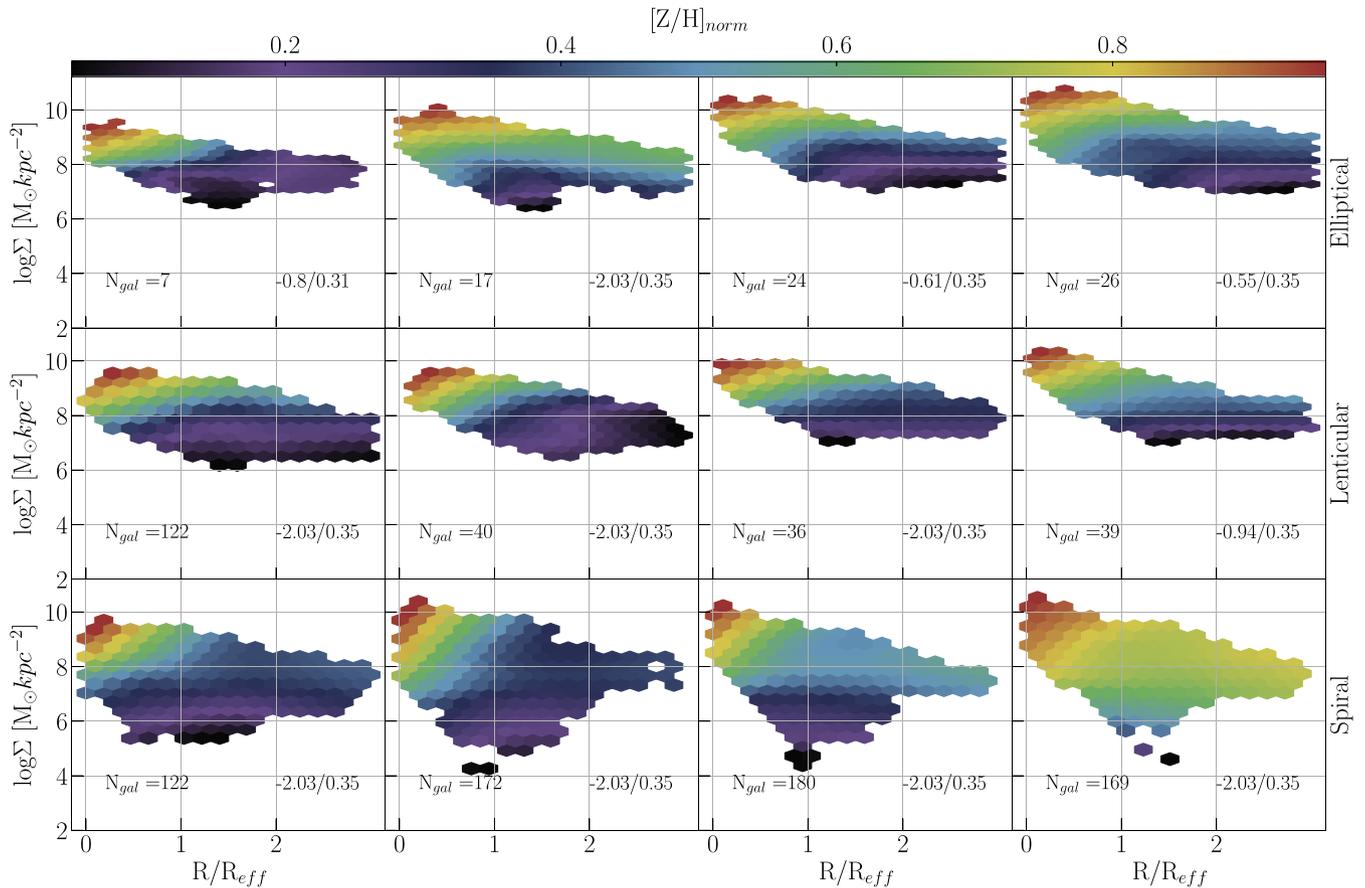

**Figure 12.** $\Sigma_*$–radius relation, colour-coded by median [Z/H] for all the spaxels in the iMaNGA sample within $3R_{\rm reff}$. The grid shows the global mass–morphology plane (columns for the stellar mass, rows for the morphology as in previous Figures and Table 1). The number of galaxies in each mass–morphology panel is reported. We use individual colour bar limits in each panel to highlight subtle trends. The minimum and maximum of each colour bar are shown in the lower right-hand corner of the corresponding panel. The data are smoothed using the LOESS algorithm.

As in N21 for MaNGA galaxies, here we find that, at any given stellar mass and morphology, the metallicity increases with the surface mass density at almost any galactocentric distance. This is expected for the iMaNGA sample given the results presented in Figs 7 and 9, and this result is consistent with what is observed in MaNGA.

As already mentioned, N21 find a constant metallicity or an *increase in metallicity with increasing radius* at a fixed stellar surface mass density for almost any morphology and any total stellar mass. Specifically, this trend is seen for low- and intermediate-mass galaxies, and it breaks for high-mass galaxies, in particular at the low $\Sigma_*$ regime. This is not seen in the iMaNGA data, as demonstrated by Fig. 12. At fixed stellar surface mass density at any point in the mass-morphology plane, *metallicity decreases (or remains constant) with increasing radius*. This anticorrelation is strongest in spiral galaxies and is in stark contrast to what is observed with MaNGA.

The conclusion in N21 is that metallicity is globally driven by galaxy mass and morphology, and locally by surface mass density. This is also what we see in the iMaNGA sample. Indeed, in both samples, we globally observe higher metallicity for more massive galaxies and for ETGs, and a local correlation between the surface mass density and the stellar metallicity. However, in both MaNGA and iMaNGA, there is evidence for galaxy radius being an additional, secondary local driver of metallicity. Interestingly, observations and simulations show opposite trends, though, with metallicity increasing with radius in MaNGA and decreasing with radius in iMaNGA at fixed surface mass density.

Furthermore, for low- and intermediate-mass galaxies in MaNGA, the [Z/H] radial profiles are flatter or even positive, while the $\Sigma_*$ radial profiles have steep negative slopes over the entire morphology–mass plane (see Figs 7 and 9 and the discussion thereby).

### 4.4 Dependence on galaxy mass and environment

Goddard et al. (2016, 2017) investigate the correlations between radial metallicity gradients, galaxy mass and environmental density. Here, we repeat the analysis for both the MaNGA catalogue and the iMaNGA sample. Total stellar mass is adopted from FIREFLY, the metallicity gradients are based on light-weighted metallicities measured within $1.5\,R_e$.

For the iMaNGA sample, we consider the environment as defined in Paper I (see section 3.1), and the gradients calculated in Section 4.2. For MaNGA, we compute the gradients in the same manner, using the information provided by the FIREFLY VAC. To associate an environment to MaNGA galaxies, we make use of the Galaxy Environment for MaNGA VAC.[5] The GEMA VAC provides galaxy environmental densities based on the Nth nearest neighbour method for 3287 MaNGA galaxies. We use this subsample of MaNGA galaxies for the following analysis. The division between ETGs and LTGs in this sample is as in Section 4.1, so using the definition adopted by N21.

---

[5]GEMA VAC https://www.sdss4.org/dr15/data_access/value-added-catalogs/

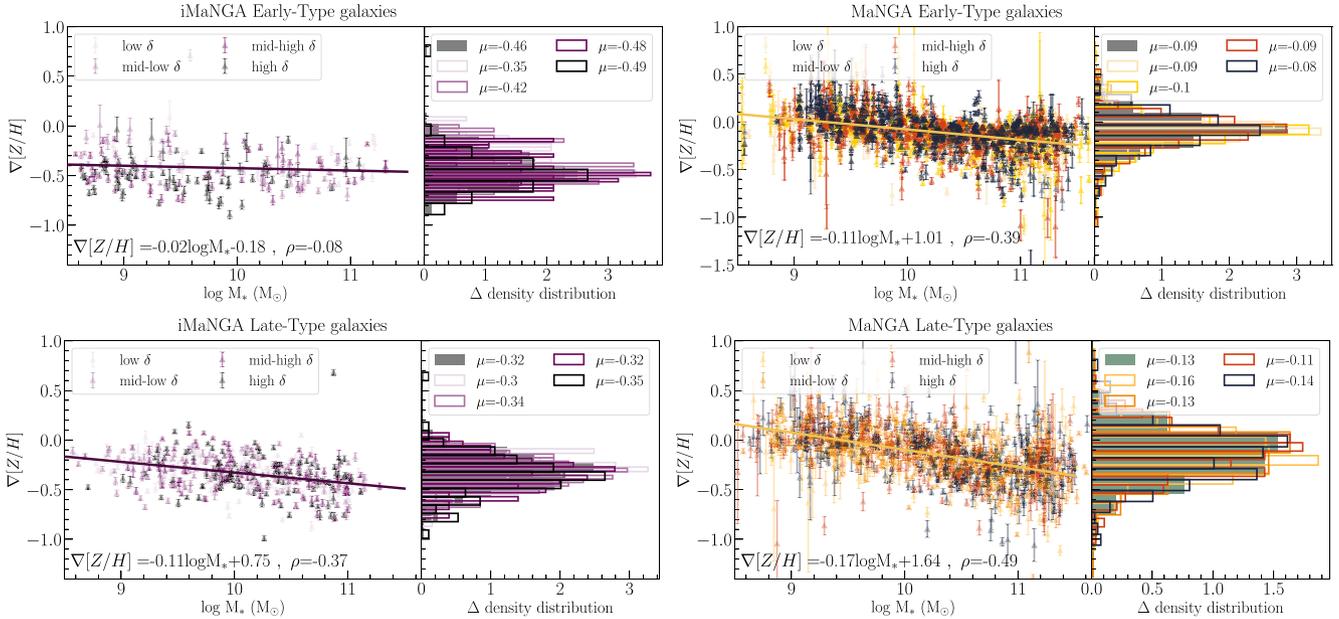

**Figure 13.** Light-weighted stellar population metallicity gradients as a function of the galaxy stellar mass, colour-coded by different local environmental densities, (see Paper I), and gradients distributions (histograms), for the iMaNGA (in violets, left column) and the MaNGA (in reds, right column) samples. The galaxies are divided into early-type (i.e. E+S0, in the upper row), and late-type galaxies (i.e. LTG in the bottom row). We report the linear regressions, the gradients and the Pearson coefficient $\rho$, considering all galaxies. For the distributions of the gradients (colour-coded by the environmental density), the median $\mu$ is reported in the legend. We also report the distributions for the entire sample, as well as the median $\mu$ (in grey).

Fig. 13 shows metallicity gradient as a function of galaxy mass for different environmental densities. Early-type galaxies are shown in the top row, late-type galaxies in the bottom row. The iMaNGA and MaNGA samples are shown by the left-hand and right-hand panels, respectively.

### 4.4.1 Galaxy mass

The figure demonstrates that the stellar metallicity gradients in early-type galaxies are systematically lower in the iMaNGA sample compared to the MaNGA sample (as already noted in Fig. 7). Most interestingly, the MaNGA data show a significant negative correlation between galaxy mass and metallicity gradient. Metallicity gradients in early-type galaxies are positive in low-mass galaxies and progressively steepen with increasing galaxy mass leading to the well-known negative gradients in intermediate- and high-mass galaxies. This pattern is not recovered by the simulations. The early-type galaxies in iMaNGA show no correlation between metallicity gradient and galaxy mass, and metallicity gradients are negative at all masses (see the Pearson coefficients reported in the figure).

This behaviour is a further manifestation of the role of surface mass density as principle local driver of stellar metallicity. The radial $\Sigma_*$ gradients presented in Fig. 12 steepen with increasing galaxy mass in MaNGA, but remain constant in iMaNGA. This discrepancy leads to the different mass dependencies of the metallicity gradients in MaNGA and iMaNGA.

The picture is somewhat different for late-type galaxies (bottom row in the figure). The metallicity gradients are again slightly steeper in iMaNGA than in MaNGA, but the steepening of the gradient with increasing galaxy mass observed in MaNGA is recovered by the simulations. The trend is, however, stronger in the observations. This is consistent with Fig. 12, in which we show that a steepening of the $\Sigma_*$ gradients with increasing mass is seen in both MaNGA and iMaNGA. Finally, it is worth noting that the stellar metallicity gradients are negative across all galaxy masses in iMaNGA, but positive in low-mass galaxies in MaNGA.

### 4.4.2 Environmental density

In Fig. 13, we colour code the galaxies in the stellar mass–gradient plane according to their environment and report the gradient distributions divided into the 4 different environments. The labels show the median of the gradient distributions $\mu$. We can see that there is no sign of any significant dependency on the environment in both MaNGA and iMaNGA for both morphological types.

This is in line with the study by Goddard et al. (2016, 2017), albeit based on a smaller sample of MaNGA galaxies. It is interesting that the observed lack of environmental dependence is replicated by the iMaNGA sample. However, the volume of the TNG50 simulations is relatively small, and it will be important to repeat this test with future simulations at the resolution of TNG50 but for larger cosmological volumes.

## 5 DISCUSSION AND CONCLUSION

We conduct a statistically and methodologically significant comparison between the MaNGA survey and the cosmological simulations TNG50. To this end, we employ a forward-modelling approach to generate a mock MaNGA sample from IllustrisTNG50, called iMaNGA, the characteristics of which are as close as possible to the observed MaNGA catalogue. In the first paper of the iMaNGA project (Paper I), we introduce our method to generate mock SDSS-IV/MaNGA integral-field spectroscopic galaxy observations from TNG50. In Paper II, we present the construction of the iMaNGA sample (see Sections 3.3 and 3.2), which we extend in this work to include the selection criteria of the MaNGA Secondary Sample yielding a final catalogue of 1500 mock MaNGA galaxies.

Following the sample selection from TNG50, we apply post-processing methods to transfer the theoretical galaxies from the simulated space into the MaNGA observational plane (see Paper I, summarized in Section 3.2). The resulting MaNGA-like data cubes are then analysed through full spectral fitting with the codes PPXF and FIREFLY adopting MaStar stellar population models following the approach of the MaNGA VAC by N22VAC.

The key of our analysis is that observational biases plaguing the interpretation of MaNGA data are emulated in the theoretical iMaNGA sample. Focusing on stellar population properties, we carry out the same analysis of the MaNGA and the iMaNGA samples. The scientific analysis discussed here follows earlier studies of MaNGA galaxies by our group presented in G17 and N21. In particular with this work, we investigate the interplay between galaxy morphology, stellar mass, stellar metallicity, stellar surface mass density, galactocentric distance, and environmental density. In the following, we discuss the main findings of this paper.

### 5.1 Stellar population scaling relations

Looking at the global mass–metallicity and –age relations in iMaNGA (Section 4.1), we show that galaxies in TNG50 recover the global trends observed in MaNGA. Indeed, ETGs are generally populated by older and more metal-rich stellar populations compared to LTGs in both samples. Both stellar age and metallicity increase with stellar mass, with iMaNGA and MaNGA following similar trends. The only significant discrepancy can be seen for light-weighted ages of lower-mass late-type galaxies with the simulations overestimating ages by about 2 Gyr. Furthermore, simulations slightly underestimate the metallicities of the most massive galaxies by about 0.1 dex. Moreover, in iMaNGA the difference between the trends followed by ETGs and LTGs is more subtle.

### 5.2 Radial profiles

In Section 4.2, we present the radial profiles for stellar age, stellar metallicity, and stellar surface mass density. We further calculate the gradients in the mass–morphology plane, as defined in Table 1, for all spaxels in the samples up to $1.5R_{\rm eff}$.

Overall, stellar metallicities are lower and metallicity gradients are steeper in iMaNGA compared to observations. We note that iMaNGA and MaNGA agree well for elliptical galaxies (except for the lowest-mass bin), with iMaNGA reproducing well the metallicity distribution in this type of galaxy. The picture is different for lenticular and spiral galaxies, for which the simulations predict significantly steeper radial metallicity gradients at any mass bin. As already found in N21, the metallicity profiles at low- and intermediate-masses ($M_* \leq 10^{10}\,{\rm M_\odot}$) become flat or even positive in MaNGA, particularly in spiral galaxies. This effect is not recovered by iMaNGA, characterized by only negative gradients.

The simulations predict higher light-weighted stellar ages at all radii. The discrepancy is as high as 4 Gyr for the lowest-mass ellipticals. However, the age gradients are consistent, except we find positive age gradients in low-mass lenticular iMaNGA galaxies, which is not seen in MaNGA. The largest discrepancy between iMaNGA and MaNGA can again be seen in spiral galaxies, mirroring the metallicity profiles in spiral galaxies, where we see significantly steeper age gradients in iMaNGA compared to MaNGA.

We also present radial profiles of the stellar mass density (see Section 4.2.3) for both iMaNGA and MaNGA. Both samples show a decrease of $\Sigma_*$ going from the centre to the outskirts of the galaxies, at any stellar mass and at any morphology. The iMaNGA sample recovers the observed $\Sigma_*$ radial profiles fairly well in massive galaxies. However, the profiles are notably steeper than the observed ones in intermediate- and low-mass galaxies – see the discussion in Section 4.2.3. This suggests low-mass objects in iMaNGA to be more compact compared to MaNGA. Interestingly, this mismatch has been identified in Paper II, where an offset in the galaxy angular sizes between the two catalogues is shown. In particular, iMaNGA is characterized by smaller galaxy angular sizes compared to MaNGA in low-mass galaxies, while the angular sizes of massive objects match (see fig. 10 in Paper II).

### 5.3 Metallicity in the radius–surface mass density plane

We also investigate the local relation between $\Sigma_*$ and stellar metallicity (see Section 4.3). Although both MaNGA and iMaNGA have a positive trend between $\Sigma_*$ and [Z/H], the trend is steeper in MaNGA. Furthermore, the simulations systematically predict lower stellar metallicities by almost a factor of 2 (0.25 dex) across all surface mass densities.

Analysing the [Z/H]–$\Sigma_*$ trends in the mass–morphology plane (Fig. 11), we find that in iMaNGA, we always have a linear positive increase between these two quantities, while in MaNGA, for low-mass galaxies, there is no clear trend. Indeed, in MaNGA, the stellar metallicity increases again toward the lowest $\Sigma_*$ values for galaxies with stellar mass $\leq 10^{10}\,{\rm M_\odot}$ and this effect is not displayed by iMaNGA. Also, in both samples, the correlation is stronger at higher mass and going from LTGs to ellipticals, and it is overall stronger in iMaNGA (as shown by the Pearson coefficient reported in the figure).

N21 find a constant metallicity or even an *increase in metallicity with increasing radius* at a fixed stellar surface mass density for almost any morphology and any total stellar mass. This is not seen in the iMaNGA data. At fixed stellar surface mass density at any point in the mass–morphology plane, *metallicity decreases (or remains constant) with increasing radius*. These results indicate the presence of a strong local correlation between the surface mass density and the stellar metallicity in both observations and simulations. In both MaNGA and iMaNGA, there is evidence for galaxy radius being an additional, secondary local driver of metallicity. Interestingly, however, observations and simulations show opposite trends, with metallicity increasing with radius in MaNGA and decreasing with radius in iMaNGA at fixed surface mass density.

### 5.4 Metallicity gradient as a function of mass and environment

Using the iMaNGA sample we also repeat the analysis presented in Goddard et al. (2016) and GD17 where the interplay between galaxy stellar mass, galaxy environmental density and metallicity gradients is investigated. In agreement with GD17, we find a significant negative correlation between metallicity gradient and the stellar mass in MaNGA, this correlation being stronger for LTGs than for ETGs. In other words, metallicity gradient gets steeper with increasing galaxy mass. This correlation is reasonably well reproduced by the simulations for LTGs, but not for ETGs. No correlation is found between metallicity gradient and stellar mass for early-type galaxies in iMaNGA. We discuss that this discrepancy is mostly caused by the lack of a mass-dependence of the gradient in surface mass density in iMaNGA.

Interestingly, both MaNGA and iMaNGA show no significant dependence of metallicity gradients on environmental density (see Section 4.4).

### 5.5 Drivers of stellar metallicity

To explain the observed trends in MaNGA, N21 propose the presence of supplementary drivers of metallicity, which, acting together with the stellar mass, enrich the stellar composition in the outskirts of the galaxies, in particular for low- and intermediate-mass galaxies ($M_*$ $\leq 10^{10.8}$ M$_\odot$). This is not seen in the simulations.

We conclude that TNG50 includes the main drivers of stellar metallicity fairly well in massive elliptical galaxies, while the interplay between stellar surface mass density, stellar metallicity and galactic distance is not fully captured for lenticular and spiral galaxies, as well as low-mass ellipticals.

An analysis of the merger history, gas metallicity, *ex situ* and *in situ* stellar populations, SMBH activity at any mass in the TNG50 galaxies here adopted might shed light on the way the local metallicity trends are built in the simulations to understand why this discrepancy, noted in particular at low and intermediate mass ($\leq 10^{10}$ M$_\odot$), arises. We can speculate that the subgrid models, such as SN and stellar feedback, have a higher impact on lower mass galaxies, and such models might not be able to fully capture the galaxy properties observed by the MaNGA survey.

Since the simulations show steeper metallicity gradients than observed, it might be important to note how galactic winds can redistribute the metals within the galaxies. The galactic winds in the subgrid models are dependent on the definition of many properties and parameters, such as wind energy, velocity, mass loading, metal loading, and/or recoupling. Changing any or more of them can significantly alter the way galactic winds act on the simulated galaxies. In particular, in the Auriga simulations (Grand et al. 2017), changing the wind metal loading factor has produced flatter metallicity gradients (Grand et al. 2019).

Further exploration and discussion of the effects of subgrid physics in TNG50 simulations will be needed to fully address the discrepancies identified in this paper between the theoretical iMaNGA sample and MaNGA observations.

### ACKNOWLEDGEMENTS

LN is supported by an STFC studentship. STFC is acknowledged for support through the Consolidated Grant Cosmology and Astrophysics at Portsmouth, ST/S000550/1. Numerical computations were done on the Sciama High Performance Compute (HPC) cluster which is supported by the ICG, SEPnet and the University of Portsmouth. JN acknowledges funding from the European Research Council (ERC) under the European Union's Horizon 2020 research and innovation programme (grant agreement no. 694343). Funding for the Sloan Digital Sky Survey IV has been provided by the Alfred P. Sloan Foundation, the U.S. Department of Energy Office of Science, and the Participating Institutions. SDSS-IV acknowledges support and resources from the Center for High Performance Computing at the University of Utah. The SDSS website is www.sdss.org. SDSS-IV is managed by the Astrophysical Research Consortium for the Participating Institutions of the SDSS Collaboration including the Brazilian Participation Group, the Carnegie Institution for Science, Carnegie Mellon University, Center for Astrophysics| Harvard & Smithsonian, the Chilean Participation Group, the French Participation Group, Instituto de Astrofísica de Canarias, The Johns Hopkins University, Kavli Institute for the Physics and Mathematics of the Universe (IPMU)/University of Tokyo, the Korean Participation Group, Lawrence Berkeley National Laboratory, Leibniz Institut für Astrophysik Potsdam (AIP), Max-Planck-Institut für Astronomie (MPIA Heidelberg), Max-Planck-Institut für Astrophysik (MPA Garching), Max-Planck-Institut für Extraterrestrische Physik (MPE), National Astronomical Observatories of China, New Mexico State University, New York University, University of Notre Dame, Observatário Nacional/MCTI, The Ohio State University, Pennsylvania State University, Shanghai Astronomical Observatory, United Kingdom Participation Group, Universidad Nacional Autónoma de México, University of Arizona, University of Colorado Boulder, University of Oxford, University of Portsmouth, University of Utah, University of Virginia, University of Washington, University of Wisconsin, Vanderbilt University, and Yale University. The primary TNG simulations were realised with compute time granted by the Gauss Centre for Supercomputing (GCS): TNG50 under GCS Large-Scale Project GCS-DWAR (2016; PIs Nelson/Pillepich), and TNG100 and TNG300 under GCS-ILLU (2014; PI Springel) on the GCS share of the supercomputer Hazel Hen at the High Performance Computing Center Stuttgart (HLRS).

### DATA AVAILABILITY

The iMaNGA catalogue is available through the following website: https://www.tng-project.org/data/docs/specifications/. All data for the analysis in this paper are hosted on the same website as iMaNGA-VAC. Finally, the IMASTAR code can be found here: https://github.com/lonanni/iMaNGA. The codes for the analysis presented in this paper, and codes related to the iMaNGA projects, are hosted on the same page.

MaNGA data are part of SDSS-IV, publicly available at (Abdurro'uf et al. 2022). The FIREFLY code is available at: https://www.icg.port.ac.uk/FIREFLY and the MaStar population models at https://www.icg.port.ac.uk/mastar. Illustris and IllustrisTNG data are publicly available at https://www.illustris-project.org/data (Nelson et al. 2019a).

### REFERENCES

Abazajian K. et al., 2003, AJ, 126, 2081
Abdurro'uf et al., 2021, ApJS, 259, 39
Abdurro'uf et al., 2022, ApJS, 259, 39
Abril-Melgarejo V. et al., 2021, A&A, 647, A152
Ade P. A. R. et al., 2016, A&A, 594, A13
Baes M., Camps P., 2015, Astron. Comput., 12, 33
Baes M., Verstappen J., De Looze I., Fritz J., Saftly W., Vidal Pérez E., Stalevski M., Valcke S., 2011, ApJS, 196, 22
Barrientos Acevedo D. et al., 2023, MNRAS, 524, 907
Blanton M. R. et al., 2017, ApJ, 154, 28
Bottrell C., Torrey P., Simard L., Ellison S. L., 2017, MNRAS, 467, 2879
Bundy K. et al., 2015, ApJ, 798, 7
Cannarozzo C. et al., 2022, MNRAS, 520, 5651
Cappellari M., 2017, MNRAS, 466, 798
Cappellari M., Copin Y., 2003, MNRAS, 342, 345
Cappellari M. et al., 2013, MNRAS, 432, 1862
Cook B. A., Conroy C., Pillepich A., Rodriguez-Gomez V., Hernquist L., 2016, ApJ, 833, 158
Davé R., Anglés-Alcázar D., Narayanan D., Li Q., Rafieferantsoa M. H., Appleby S., 2019, MNRAS, 486, 2827
Dawson K. S. et al., 2013, AJ, 145, 10

De Rossi M. E., Bower R. G., Font A. S., Schaye J., Theuns T., 2017, MNRAS, 472, 3354
Dolag K., 2015, in IAU General Assembly. p. 2250156
Domínguez Sánchez H., Margalef B., Bernardi M., Huertas-Company M., 2022, MNRAS, 509, 4024
Drory N. et al., 2015, AJ, 149, 77
Feldmann R. et al., 2023, MNRAS, 522, 3831
Gallazzi A., Charlot S., Brinchmann J., White S. D. M., Tremonti C. A., 2005, MNRAS, 362, 41
Genel S. et al., 2014, MNRAS, 445, 175
Goddard D. et al., 2016, MNRAS, 466, 4731
Goddard D. et al., 2017, MNRAS, 466, 4731
Grand R. J. J. et al., 2017, MNRAS, 467, 179
Grand R. J. J. et al., 2019, MNRAS, 490, 4786
Groves B., Dopita M. A., Sutherland R. S., Kewley L. J., Fischera J., Leitherer C., Brandl B., van Breugel W., 2008, ApJS, 176, 438
Gunn J. E. et al., 2006, AJ, 131, 2332
Hill L. et al., 2021, MNRAS, 509, 4308
Hubble E., 1926, Contrib. Mount Wilson Obs./Carnegie Inst. Wash., 324, 1
Huertas-Company M. et al., 2019, MNRAS, 489, 1859
Kaviraj S. et al., 2017, MNRAS, 467, 4739
Kroupa P., 2002, Science, 295, 82
Law D. R. et al., 2016, AJ, 152, 83
Lian J. et al., 2018, MNRAS, 476, 3883
Maraston C., 2005, MNRAS, 362, 799
Maraston C., Strömbäck G., 2011, MNRAS, 418, 2785
Maraston C. et al., 2020, MNRAS, 496, 2962
Marinacci F. et al., 2018, MNRAS
McAlpine S. et al., 2016, Astron. Comput., 15, 72
McElwain M. W. et al., 2023, Publ. Astron. Soc. Pac., 135, 34
Naiman J. P. et al., 2018, MNRAS, 477, 1206
Nanni L. et al., 2022, MNRAS, 515, 320
Nanni L. et al., 2023, MNRAS, 522, 5479
Nelson D. et al., 2018, MNRAS, 475, 624
Nelson D. et al., 2019a, Comput. Astrophys. Cosmology, 6, 2
Nelson D. et al., 2019b, MNRAS, 490, 3234
Neumann J. et al., 2021, MNRAS, 508, 4844
Neumann J. et al., 2022, MNRAS, 513, 5988
Oyarzún G. A. et al., 2019, ApJ, 880, 111
Petrosian V., 1976, ApJ, 210, L53
Pillepich A. et al., 2018a, MNRAS, 473, 4077
Pillepich A. et al., 2018b, MNRAS, 475, 648
Pillepich A. et al., 2019, MNRAS, 490, 3196
Rodriguez-Gomez V. et al., 2019, MNRAS, 483, 4140
Ryden B. S., 2004, ApJ, 601, 214
Salpeter E. E., 1955, ApJ, 121, 161
Sánchez S. F. et al., 2016, RMxAA, 52, 21
Sánchez S. F. et al., 2022, ApJS, 262, 36
Sarmiento R., Huertas-Company M., Knapen J. H., Ibarra-Medel H., Pillepich A., Sánchez S. F., Boecker A., 2022, A&A, 673, 22
Schaye J. et al., 2014, MNRAS, 446, 521
Schulz S., Popping G., Pillepich A., Nelson D., Vogelsberger M., Marinacci F., Hernquist L., 2020, MNRAS, 497, 4773
Sérsic J. L., 1963, Bol. Asociacion Argentina Astron. La Plata Argentina, 6, 41
Sersic J. L., 1968, Observatorio Astronomico, Universidad Nacional de Cordoba. Argentina, Atlas de Galaxias Australes
Sijacki D., Vogelsberger M., Genel S., Springel V., Torrey P., Snyder G. F., Nelson D., Hernquist L., 2015, MNRAS, 452, 575
Smee S. A. et al., 2013, AJ, 146, 32
Springel V. et al., 2018, MNRAS, 475, 676
Sánchez-Blázquez P. et al., 2006, MNRAS, 371, 703
Taylor P., Kobayashi C., 2014, MNRAS, 442, 2751
Tonini C., Maraston C., Thomas D., Devriendt J., Silk J., 2010, MNRAS, 403, 1749
Torrey P. et al., 2015, MNRAS, 447, 2753
Torrey P. et al., 2019, MNRAS, 484, 5587
Trayford J. W. et al., 2015, MNRAS, 452, 2879
Trayford J. W. et al., 2017, MNRAS, 470, 771
Tremonti C. A. et al., 2004, ApJ, 613, 898
Villaescusa-Navarro F. et al., 2021, ApJ, 915, 71
Vogelsberger M. et al., 2014, MNRAS, 444, 1518
Wake D. A. et al., 2017a, AJ, 154, 86
Wake D. A. et al., 2017b, AJ, 154, 86
Weinberger R. et al., 2017, MNRAS, 465, 3291
Westfall K. B. et al., 2019, AJ, 158, 231
Wilkinson D. M., Maraston C., Goddard D., Thomas D., Parikh T., 2017, MNRAS, 472, 4297
Yan R. et al., 2019, ApJ, 883, 175
York D. G. et al., 2000, AJ, 120, 1579

## APPENDIX A: TOTAL STELLAR MASS IN MANGA AND IMANGA

Fig. A1 illustrates the relation between different definitions of the total stellar mass, for both the iMaNGA and MaNGA samples. In particular, on the $x$-axis, we report the total stellar mass, as obtained by running the full-spectral fitting code FIREFLY, for both MaNGA and iMaNGA. On the $x$-axis, instead, for the MaNGA galaxies, we report the total stellar mass as provided by NSA catalogue, i.e. based on the photometry. For the iMaNGA catalogue, instead, the total stellar mass is defined as the sum of the stellar masses of any stellar particles in the respective TNG50 galaxy, as defined by the SUBFIND algorithm (see Section 2.2). We can see there is a tight-linear correlation between these definitions of total stellar mass and the two catalogues follow similar trends. To enhance the consistency in the discussion within the paper, we used the total stellar mass computed considering the FIREFLY information for both catalogues, also because the SUBFIND algorithm for the TNG50 galaxies can associate with the galaxies stellar particles at a distance from the centre of mass of the galaxy of above 100 kpc (see for example Fig. 2 in Paper I).

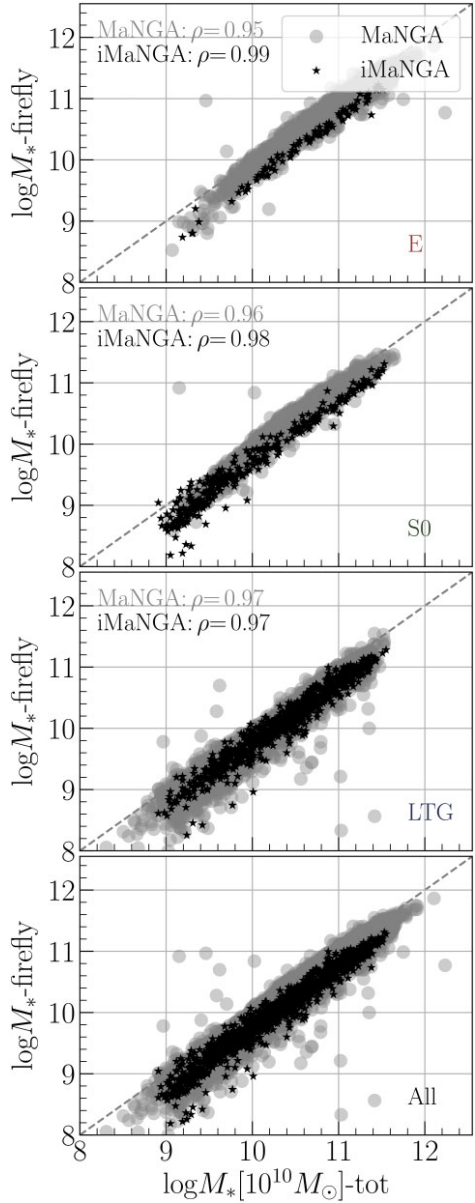

**Figure A1.** Stellar mass computed by full-spectral fitting with FIREFLY, for both the MaNGA (grey circles) and iMaNGA (black stars) sample, as a function of the total stellar mass, assuming the information provided by TNG50 for the iMaNGA sample and the NSA photometry-based stellar mass for the MaNGA sample, for the entire sample (bottom panel) and dividing the samples in morphology (first three panels). For consistency in the analysis, the stellar mass computed by FIREFLY is used for both samples throughout the paper. The Pearson correlation coefficient for both catalogues is reported in the upper right corner of each panel.

## APPENDIX B: T-MORPHOLOGY

Fig. B1 shows the relation between the Sérsic index and the total stellar mass, colour-coded by the T morphology, as associated with each galaxy through visual inspections by the authors. For simplicity, only 4 values of the T-morphology are adopted in the (Rodriguez-Gomez et al. 2019). We adopt the T-morphology instead of the Sérsic index within this work, to follow the steps in N21 as closely as possible for a fair comparison between the observed galaxies and the simulated ones.

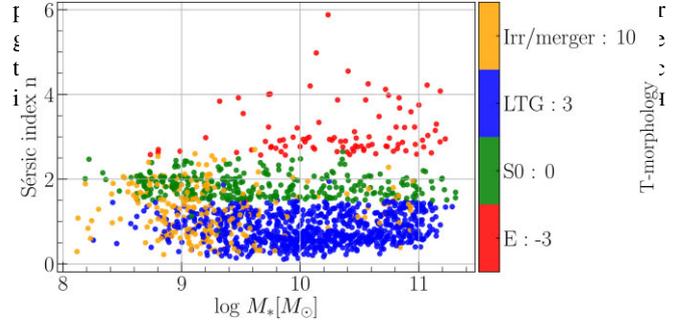

**Figure B1.** The distribution of the galaxies in the iMaNGA sample in the stellar mass-morphology plane, colour-coded by the T-morphology associated with each galaxy. The Sérsic index is computed for each galaxy in the sample with the Sérsic 2D fitting code STATMORPH (see Paper I), while the T-morphology is based on visual inspection. In particular, 4 values of T-morphology are used in this paper: 10 for irregulars or galaxies showing signs of mergers, 3 for spirals (LTG), 0 for lenticulars (S0) and −3 for ellipticals (E).

## APPENDIX C: TNG50 INTRINSIC GRADIENTS AND BIAS IN THE RECOVERED QUANTITIES

Fig. C1 shows the distribution of the bias, between 'intrinsic' and 'recovered' quantities (see Paper II) in the mass–morphology plane. In Paper II, we show the bias over the entire catalogue, here we show them in the plane, to demonstrate how, in any bin (see Table 1), the 'intrinsic' quantities in the galaxies in TNG50 are recovered by running the full-spectral fitting code FIREFLY over the mock MaNGA data cubes. It can be appreciated how with our methodology for the construction of the data cubes, and the analysis, no particular bias are introduce, and the intrinsic and recovered properties agree with zero at the 68 per cent confidence level. It is interesting to note that the biases increase going from elliptical to spiral galaxies; this can be explained by the increase of dust attenuation (see also the discussion in Paper II, where we show how the dust particles are found in galaxies which are categorized as lenticulars and spirals thanks to the Sérsic index provided by STATMOPRH in the iMaNGA catalogue) and the effects of inclination.

Fig. C2 demonstrate how the 'intrinsic' gradients in TNG50 are recovered in our methodology, i.e. the differences between MaNGA and iMaNGA are not generated by the methodology followed to create and analyse the MaNGA data cubes (see Paper I and Paper II) but intrinsic to the simulations. Here, we are presenting the results for MW quantities, to make solely use of information available in TNG50. A discussion over LW quantities for TNG50 'intrinsic' properties is presented in Paper II, where, as for MW 'intrinsic' quantities, bias in the residuals are not found.

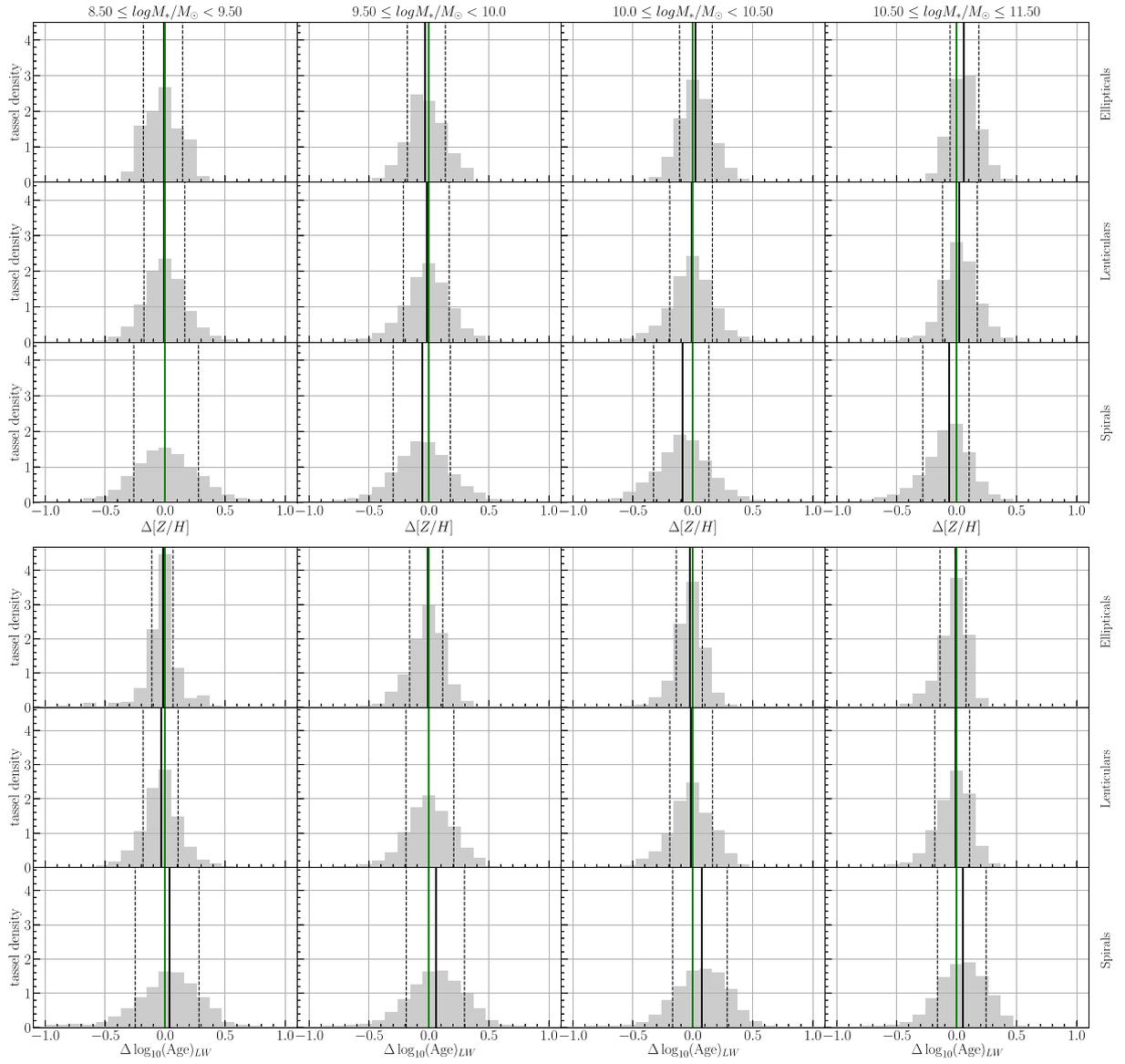

**Figure C1.** Distributions of the bias in the morphology-mass plane in the recovery of the intrinsic quantities in TNG50 running the full spectral fitting algorithm FIREFLY on the mock MaNGA-data cubes in the iMaNGA galaxy catalogue.

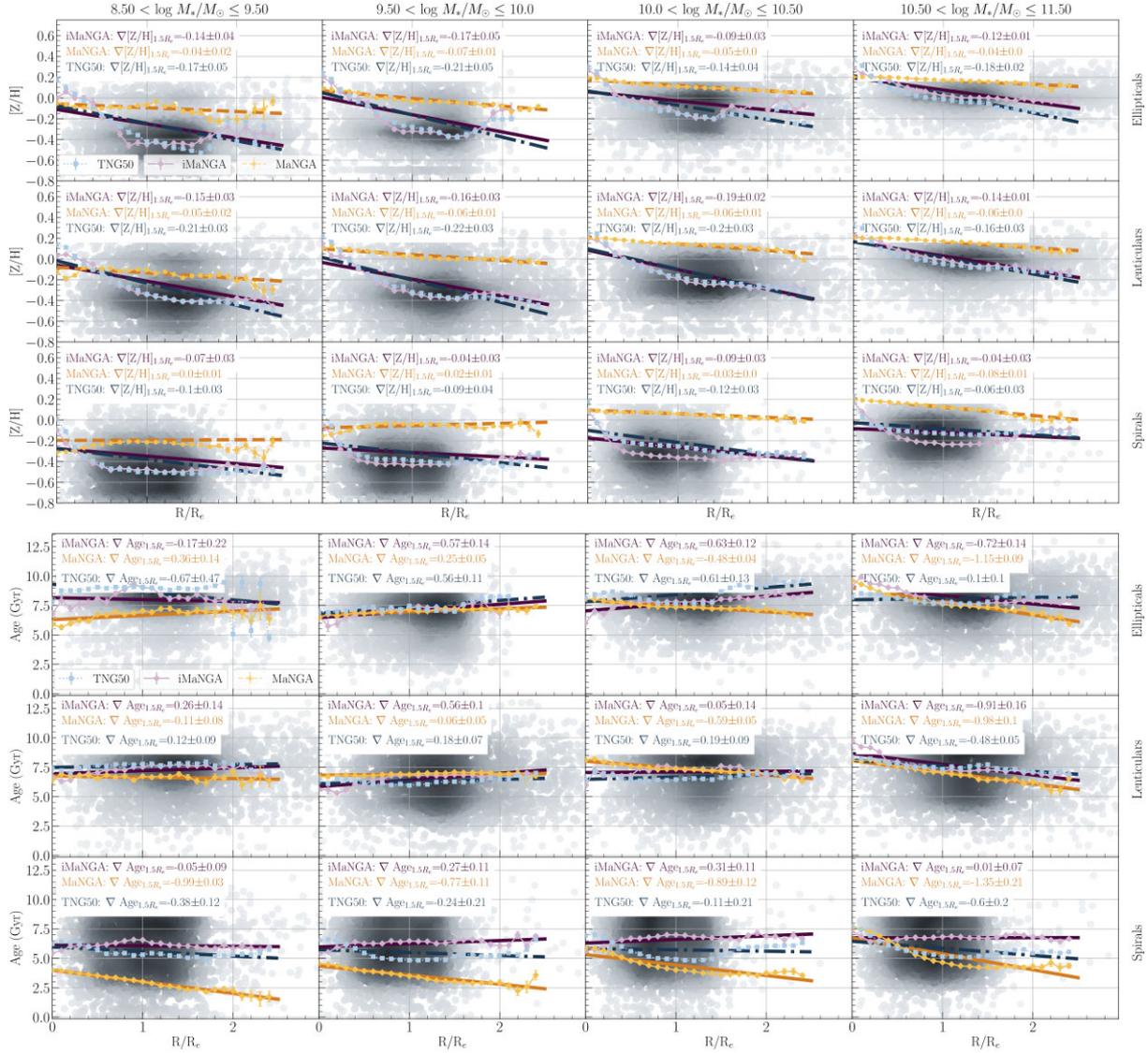

**Figure C2.** Comparison of the gradients in metallicity and age computed from the stellar populations' information for the MaNGA, iMaNGa and TNG50 galaxies. The gradients are computed as in Figs 7 and 8. For iMaNGA and MaNGA the information is shown with the same symbols and styles as in Fig. 7. Here, the TNG50 intrinsic information are reported in blues.

This paper has been typeset from a T_EX/L^AT_EX file prepared by the author.